\documentclass[aps,prx,twocolumn,superscriptaddress]{revtex4-2}
\usepackage{amsmath}
\usepackage{amsfonts}
\usepackage{amssymb}
\usepackage{dsfont}
\usepackage{enumerate}
\usepackage{graphicx}
\usepackage{mathtools}
\usepackage{xcolor}
\usepackage[colorlinks=True,citecolor=myblue,linkcolor=myorange,urlcolor=mygreen]{hyperref}
\usepackage{hypcap}
\usepackage{braket}
\usepackage{bm}
\usepackage[export]{adjustbox}
\usepackage{verbatim}
\usepackage[normalem]{ulem}

\newcommand{\ssm}{\mathsf{m}}
\newcommand{\ssC}{\mathsf{C}}
\newcommand{\ssQ}{\mathsf{Q}}
\newcommand{\ssS}{\mathsf{S}}

\definecolor{myblue}{RGB}{0,0,130}
\definecolor{myorange}{RGB}{130,50,0}
\definecolor{mygreen}{RGB}{0,130,0}

\begin{document}
\def\papertitle{Machine learning the effects of many quantum measurements}
\title{\papertitle}
\author{Wanda Hou}
\thanks{WH and SJG contributed equally to this work.}
\affiliation{Department of Physics, University of California, San Diego, La Jolla, California 92093, USA}
\author{Samuel J. Garratt}
\thanks{WH and SJG contributed equally to this work.}
\affiliation{Department of Physics, University of California, Berkeley, California 94720, USA} \affiliation{Department of Physics
Princeton University, Princeton, NJ 08544, USA}
\author{Norhan M. Eassa}
\affiliation{Google Research}
\affiliation{Department of Physics and Astronomy, Purdue University, West Lafayette, IN 47906, USA}
\author{Eliott Rosenberg}
\affiliation{Google Research}
\author{Pedram Roushan}
\affiliation{Google Research}
\author{Yi-Zhuang You}
\affiliation{Department of Physics, University of California, San Diego, La Jolla, California 92093, USA}
\author{Ehud Altman}
\affiliation{Department of Physics, University of California, Berkeley, California 94720, USA} 
\affiliation{Materials Sciences Division, Lawrence Berkeley National Laboratory, Berkeley, CA 94720, USA}


\begin{abstract}
Measurements are essential for the processing and protection of information in quantum comput-
ers. They can also induce long-range entanglement between unmeasured qubits. However, when post-measurement states depend on many non-deterministic measurement outcomes, there is a barrier to observing and using the entanglement induced by prior measurements. Here we demonstrate a new approach for detecting such measurement-induced entanglement. We create short-range entangled states of one- and two-dimensional arrays of qubits in a superconducting quantum processor, and aim to characterize the long-range entanglement induced between distant pairs of qubits when we measure all of the others. To do this we use unsupervised training of neural networks on observations to create computational models for post-measurement states and, by correlating these models with experimental data, we reveal measurement-induced entanglement. Our results additionally demonstrate a transition in the ability of a classical agent to accurately model the experimental data; this is closely related to a measurement-induced phase transition. We anticipate that our work can act as a basis for future experiments on quantum error correction and more general problems in quantum control.
\end{abstract}

\maketitle

\textbf{Introduction. ---}
A quantum computer naturally represents superpositions of $2^N$ classical bit strings using only $N$ qubits. However, in order to extract information from a quantum state we must perform measurements, and the outcomes that we observe are intrinsically random. This feature of quantum mechanics imposes fundamental limitations on computational power \cite{gross2009most}. Indeed, if it were possible to efficiently steer ourselves toward a desired measurement outcome, i.e. to postselect, then we could use quantum systems to efficiently solve computational problems that are expected to be fundamentally hard, under standard complexity theoretic assumptions \cite{abrams1998nonlinear,aaronson2005quantum,bremner2011classical}. A basic scientific problem, then, is to understand what the barrier to postselection implies for our ability to characterize quantum states. 

\begin{figure*}
\centering
\includegraphics[width=0.82\textwidth]{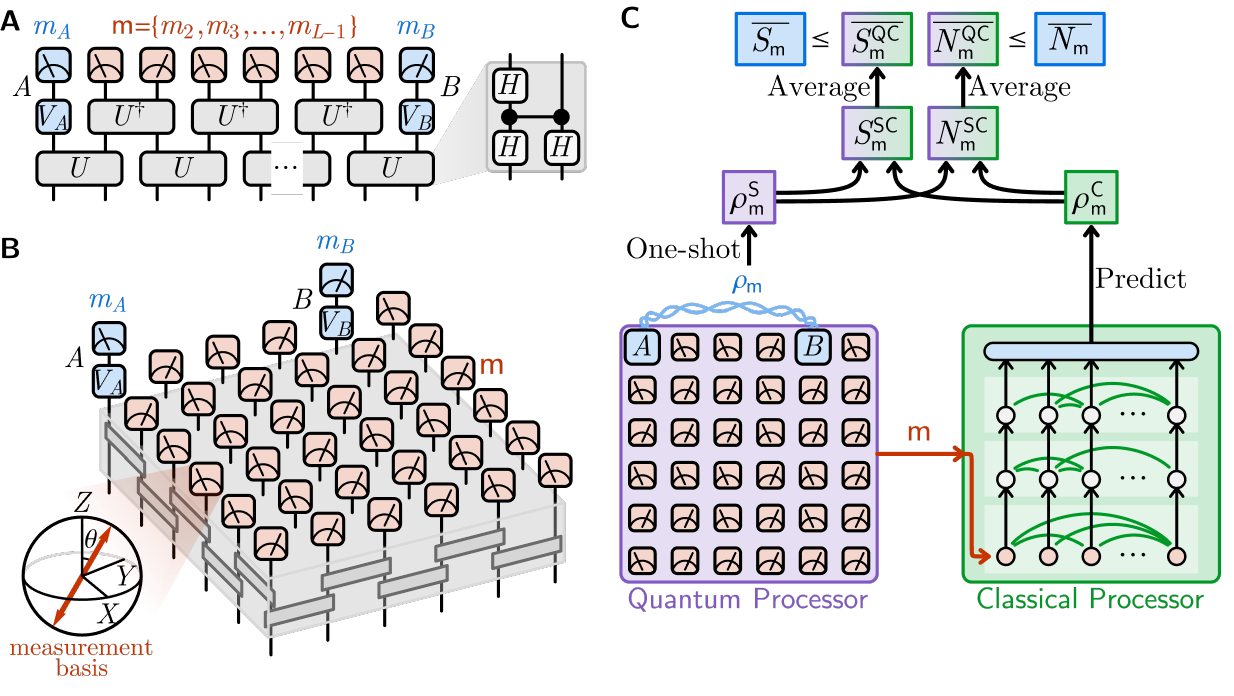}
\caption{\textsf{A}: One-dimensional cluster states. We generate cluster states using gates $U$ and $U^{\dag}$ (grey boxes), each constructed from controlled-$Z$ ($\textrm{CZ}$) and Hadamard ($H$) gates. Measuring all qubits except for the probes $A$ and $B$ (at the ends of the chain) generates entanglement between the probes. \textsf{B}: Two-dimensional cluster states. We generate cluster states in square arrays using $\textrm{CZ}$ gates between all neighboring pairs of qubits as well as single-qubit gates. Measuring all qubits except for probes, in a basis parameterized by $\theta$, leads to measurement-induced entanglement between the probes for $\theta$ near $\pi/2$, but vanishing entanglement for $\theta=0$. Increasing $\theta$ leads to a sharp onset of entanglement at $\theta = \theta_c$. In both kinds of experiments (illustrated in  $\textsf{A}$ and $\textsf{B}$) shadows $\rho^{\ssS}_{\ssm}$ of probe qubits are obtained by applying random single-qubit unitary operations and measuring. In practice these measurements are performed simultaneously with those on non-probe qubits, i.e. those used to to prepare $\rho_{\ssm}$. \textsf{C}: Unsupervised learning of post-measurement states from experimental data. Sets of outcomes $\ssm$ on non-probe qubits from the experiment (left) are used as input to an attention-based generative neural network (right), which outputs an estimate $\rho^{\ssC}_{\ssm}$ for the post-measurement state of the probe qubits $\rho_{\ssm}$. In training $\rho^{\ssC}_{\ssm}$ is combined with experimental shadows $\rho^{\ssS}_{\ssm}$ to construct a loss function, and this function is minimized to improve the prediction $\rho^{\ssC}_{\ssm}$. After training, $\rho^{\ssC}_{\ssm}$ is combined with previously unseen data in order to construct a lower bound $\overline{N^{\ssQ \ssC}_{\ssm}}$ on measurement-induced entanglement negativity $\overline{N_{\ssm}}$ between probe qubits and an upper bound $\overline{S^{\ssQ \ssC}_{\ssm}}$ on the measurement-averaged von Neumann entropy $\overline{S_{\ssm}}$ of the probes.}
\label{fig:diagram}
\end{figure*}

In this work we set out to observe the effects of large numbers of measurements on many-qubit states. The most remarkable consequence of measurement is quantum collapse, which can involve a nonlocal change in a state at an arbitrarily large distance \cite{bell1964einstein}. Because these nonlocal changes depend on non-deterministic measurement outcomes, they are only visible to interactive observers who actively use the outcomes, as in quantum teleportation \cite{bennett1993teleporting}. Indeed, if a nonlocal change in a state could be observed without such processing, entanglement would be a resource for superluminal communication. When many measurements are performed, exotic topological orders \cite{raussendorf2005long,tantivasadakarn2024long,lu2022measurement} and measurement-induced critical states \cite{skinner2019measurement,li2018quantum,bao2020theory,jian2020measurement,garratt2023measurements} can emerge from collapse, but it is \emph{a priori} unclear how interactive observers should process and use their information to detect these structures.

One possible strategy is to prepare an ensemble of identical post-measurement states. This allows the observer to directly measure expectation values or to perform quantum-state tomography within the ensemble. The key problem with this approach is that, due to the no-cloning theorem \cite{nielsen2010quantum}, the only way to prepare this ensemble is to postselect on specific sets of measurement outcomes. Because each set of outcomes occurs with a probability that is exponentially small in the number of measurements performed, this strategy requires that the experiment is repeated exponentially many times. For this reason, experiments following this strategy are limited to small systems \cite{koh2023measurement}. 

To avoid this exponential time requirement an observer can employ a computational model, which takes
observed measurement outcomes as input and generates predictions for the corresponding post-measurement states \cite{gullans2020scalable}. Given such a model of the system, it is possible to efficiently verify whether properties of the experimentally prepared post-measurement states resemble the model predictions. Such schemes have been implemented in experiments on arrays of trapped ions \cite{noel2022measurement,agrawal2024observing} and superconducting qubits \cite{google2023measurement,kamakari2025experimental} to detect measurement-induced phase transitions. Moreover, even if such models are imperfect, they can be used to bound physical properties of the experimentally prepared states, thereby providing an objective characterization of the effects of measurements \cite{garratt2024probing,mcginley2024postselection}. 

With prior knowledge of how the system was prepared, the computational model used for this task could be a direct simulation of the process. However, such information is not readily available in the experimental measurement outcomes. This raises the question of whether the effects of quantum collapse can be efficiently decoded from measurement outcomes without prior knowledge of the system. This is essentially a question about learning. When can the observer learn to predict the post-measurement state from the data, and thereby observe the nonlocal effects of quantum collapse?

To investigate this question, here we prepare cluster states \cite{raussendorf2001one} of one- and two-dimensional qubit arrays on superconducting quantum processors (Google Sycamore \cite{arute2019quantum} and Willow \cite{acharya2024quantum} devices, respectively). By performing large numbers of measurements, we attempt to induce entanglement between a pair of well-separated unmeasured `probe' qubits. The interactive observer, tasked with detecting the measurement-induced entanglement (MIE), is a generative neural network (NN). This NN is trained on experimental data, without supervision, to predict the quantum state of the probe qubits from the outcomes of measurements on all other qubits.

In the one-dimensional array, cross-correlations between the predictions of the model and the experimental data reveal MIE between the two ends of an array of $34$ qubits. Due to decoherence and measurement errors, we detect the MIE by constructing bounds on mixed-state entanglement measures, such as the entanglement negativity \cite{vidal2003computable}. For this setup we find that generative NNs, which make predictions based on measurement data alone and have no additional information about the underlying cluster state, perform just as well as a computational model based on knowledge of the gates used to prepare the state.

In the two-dimensional array, we tune the system through a finite-size version of a measurement-induced phase transition \cite{li2018quantum,skinner2019measurement} by varying the measurement basis. The entanglement between the pair of spatially separated probe qubits is expected to onset non-analytically at the critical point in an infinite system \cite{bao2024finite}. We demonstrate that the critical point (corresponding to a critical angle between measurement and computational bases) is associated with a transition in the ability of the NN to learn, and thereby observe, MIE. In the phase where the probe qubits are highly entangled, the NNs
fail to reconstruct accurate models for post-measurement states from experimental data, instead generating almost featureless predictions for post-measurement states. As a consequence, the NN fails to detect the entanglement. In the vicinity of the transition we find a peak in the entanglement negativity, and also a peak in the amount of information learned by the NN as both the amount of training data and the training time are increased. The latter are new signatures of a learnability transition~\cite{bao2020theory,dehghani2023neural,ippoliti2024learnability,agrawal2024observing}, and our results indicate that they can be detected with experimental and computational resources that are only polynomial in the number of qubits.

\textbf{Experiments. ---} Our experiments consist of the following steps. First, we prepare a short-range entangled cluster state of many qubits. Second, we measure all but two of the qubits in a specified basis. This step prepares a post-measurement state $\rho_{\ssm} = \rho_{AB,\ssm}$ of the two probe qubits $A$ and $B$, which depends on the set of outcomes $\ssm$ (a string of bits); see Fig.~\ref{fig:diagram}\textsf{A}. 
In the third step, we measure the two probe qubits in a random Pauli basis, allowing us to construct a classical shadow \cite{huang2020predicting} of the state $\rho_{\ssm}$.
Our aim is to detect entanglement between $A$ and $B$ in the set of states $\rho_{\ssm}$. However, we cannot directly use the classical shadows for this purpose, e.g., through averaging over them, because with high probability we obtain a different $\ssm$ and hence a different state $\rho_{\ssm}$ in every repeat of the experiment. As mentioned in the introduction and discussed in more detail below, we can obtain an objective estimate of the entanglement by correlating the measurements of the probe qubits with the predictions of a computational model. 

\begin{figure}
\includegraphics[width=0.47\textwidth]{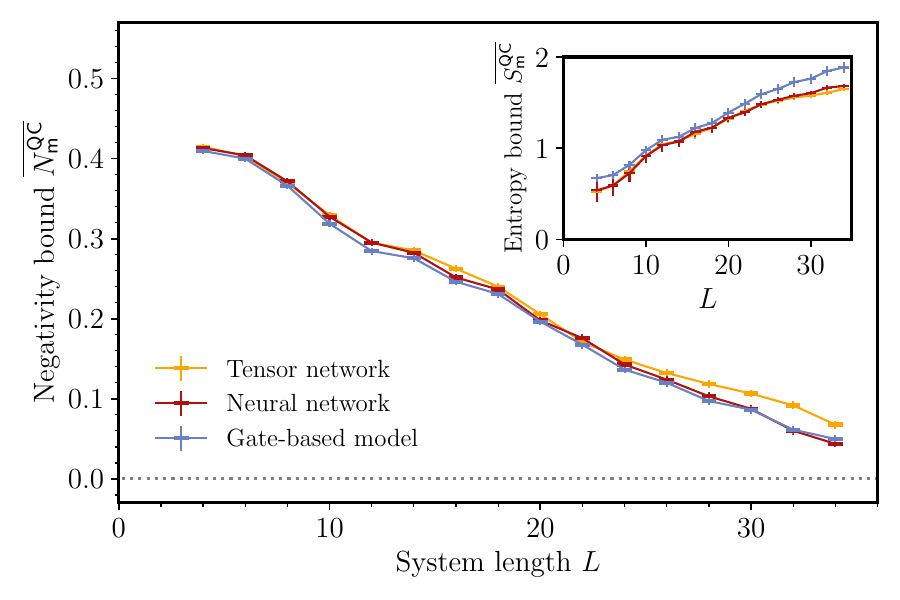}
\caption{Measurement-induced entanglement in one-dimensional cluster states. The main panel shows lower bounds $\overline{N^{\ssQ \ssC}_{\ssm}}$ on measurement-averaged negativity $\overline{N_{\ssm}}$. Different colors correspond to bounds constructed using different computational models $\ssm \mapsto \rho^{\ssC}_{\ssm}$ in Eq.~\eqref{eq:NQC}. We use two kinds of generative machine learning models, involving (red) attention mechanisms and (orange) tensor network representations of the state, and we compare these with (blue) models for post-measurement states based on knowledge of gates. The inset shows upper bounds $\overline{S^{\ssQ \ssC}_{\ssm}}$ on the von Neumann entropies $\overline{S_{\ssm}}$ of the two probe qubits, constructed using the same methods as above. The error bars here, and elsewhere, indicate the standard error in the average over repeats of the experiment.}
\label{fig:chain}
\end{figure}

In the one-dimensional geometry, our setup consists of an even number $L$ of superconducting qubits $j=1,\ldots,L$ in a line, initially prepared in a product state stabilized by all Pauli $Z_j$ operators. A cluster state is then prepared using a depth-$2$ unitary circuit composed of two layers of two-qubit gates: the first layer of gates acts on qubit pairs $(2k-1, 2k)$ for $k=1,\ldots,L/2$, transforming operators $Z_{2k-1}\mapsto Z_{2k-1} Z_{2k}$ and $Z_{2k}\mapsto X_{2k-1} X_{2k}$, and the second layer of gates acts on $(2k,2k+1)$ for $k=1,\ldots,L/2-1$ with the inverse of this transformation; see Fig.~\ref{fig:diagram}\textsf{A}. Measuring $Z_j$ on qubits $j=2,\ldots,L-1$, and finding a set of outcomes $\ssm=m_2, \ldots, m_{L-1}$, creates a state $\rho_{\ssm}$ of the probe qubits $A$ and $B$, here $j=1$ and $j=L$ respectively. The form of $\rho_{\ssm}$ depends on the set of measurement outcomes $\textsf{m}$ observed on the $L-2$ `preparation' qubits and, in the absence of noise, $\rho_{\ssm}$ is one of four (mutually orthogonal) pure maximally entangled two-qubit states. The results of our experiments on one-dimensional arrays are shown in Fig.~\ref{fig:chain}.

Following this, we study MIE arising from cluster states of two-dimensional ($6 \times 6$) qubit arrays. We prepare initial cluster states using short-depth unitary circuits having the following structure. Starting from initial states stabilized by all $Z_j$ operators, we apply (i) a Hadamard to each system qubit, and (ii) $ZZ$ gates, $\exp[i(\pi/4)Z_j Z_k]$, between all neighboring pairs of system qubits $j,k$. Measuring operators $\cos[\phi] X_j+\sin[\phi]Y_j$ in the resulting cluster state can then be viewed as inducing a measurement-based quantum computation, while measuring a $Z_j$ operator simply disentangles qubit $j$. To vary the measurement-basis between these two extremes, we apply single-qubit unitary operations $\exp[i(\theta/2) Y_j]\exp[i(\phi/2) Z_j]$ to all system qubits, with $\phi=5\pi/4$ fixed and $\theta$ variable, realizing random post-measurement states on the remaining probe qubits. Here probe qubits are separated by a distance $d$ along one edge of the square array, as illustrated in Fig.~\ref{fig:diagram}\textsf{B}. In the absence of noise, measuring $Z_j$ on all system qubits except for the probes generates a pure two-qubit state $\rho_{\ssm}$ of $A$ and $B$, with the degree of entanglement between $A$ and $B$ depending on $\ssm$ and $\theta$. For $\theta=0$ and $d>1$, $\rho_{\ssm}$ is a product state, while we expect that at $\theta=\pi/2$ the probe qubits are entangled even when they are separated by an arbitrarily large distance. 

In both kinds of experiments, in order to detect entanglement in the states $\rho_{\ssm}$, we measure the two probe qubits in random bases. This involves applying random single-qubit unitary operations $V_A$ and $V_B$, drawn independently from the set $\{\mathds{1},e^{i\frac{\pi}{4}X},e^{i\frac{\pi}{4}Y}\}$, to each of the probe qubits $A$ and $B$. Following this, we also apply a set of $\textrm{CNOT}$ gates used for measurement error detection (Methods).

After applying all of the unitary operations described above, we simultaneously measure all system and error detection qubits in the $Z$ basis. Note that, even though our aim is to diagnose the effects of measurements of non-probe system qubits on the states of probe qubits, we can perform all measurements at the same time because the local operators that we measure commute. The distinction between the measurements which generate $\rho_{\ssm}$ and the measurements used to probe $\rho_{\ssm}$ arises in classical post-processing.

From the random single-qubit unitary operations $V_A$ and $V_B$, and the outcomes $m_A$ and $m_B$ observed on the probe qubits, we construct a classical shadow \cite{huang2020predicting} of $\rho_{\ssm}$, and we denote this by $\rho^{\ssS}_{\ssm}$. For brevity, we label these shadows by the set of outcomes $\ssm$ observed on the non-probe qubits. The shadows are given by $\rho^{\ssS}_{\ssm} = \rho^{\ssS}_{\ssm,A}\otimes \rho^{\ssS}_{\ssm,B}$, where $\rho^{\ssS}_{\ssm,A} = 3 \ket{\psi_{\ssm,A}}\bra{\psi_{\ssm,A}}-\mathds{1}$ with $\ket{\psi_{\ssm,A}}=V_A^{\dag}\ket{m_A}$ and $\ket{\psi_{\ssm,B}}=V_B^{\dag}\ket{m_B}$. We then use the experimental data, which consist of a set of outcomes $\ssm$ and a single two-qubit shadow $\rho^{\ssS}_{\ssm}$ collected from each repeat of the experiment, along with the framework discussed in the next section, to detect entanglement in the ensemble of post-measurement states $\rho_{\ssm}$. 

\textbf{Cross-correlations. ---}
Given a computational model for the post-measurement density matrix, which is a function $\ssm \mapsto \rho^{\ssC}_{\ssm}$ whose input is the set $\ssm$ of outcomes and whose output is an estimate $\rho^{\ssC}_{\ssm}$ for $\rho_{\ssm}$, we can bound the true measurement-averaged entanglement in the ensemble of physical states $\rho_{\ssm}$~\cite{garratt2024probing}. The quantum-classical von Neumann entropy $S^{\ssQ \ssC}_{\ssm}$ is defined by $S^{\ssQ \ssC}_{\ssm} = -\text{Tr}[\rho_{\ssm}\log_2\rho^{\ssC}_{\ssm}]$, and can be expressed as $S^{\ssQ \ssC}_{\ssm}=S_{\ssm}+D^{\text{KL}}_{\ssm}$ where $S_{\ssm} = -\text{Tr}[\rho_{\ssm}\log_2\rho_{\ssm}]$ is the true von Neumann entropy of $\rho_{\ssm}$ and $D^{\text{KL}}_{\ssm}$ is the quantum relative entropy, or quantum Kullback-Leibler (KL) divergence, between $\rho^{\ssC}_{\ssm}$ and $\rho_{\ssm}$. Since $D^{\text{KL}}_{\ssm}\geq0$ we have $S^{\ssQ \ssC}_{\ssm} \geq  S_{\ssm}$. Crucially, from the cross-correlation between the shadows and our computational model, we can determine the average over observed outcomes $\ssm$ of $S^{\ssS \ssC}_{\ssm} \equiv -\text{Tr}[\rho^{\ssS}_{\ssm}\log_2\rho^{\ssC}_{\ssm}]$, giving \cite{garratt2024probing}
\begin{align}
    \overline{S^{\ssS \ssC}_{\ssm}} = \overline{S^{\ssQ \ssC}_{\ssm}} \geq  \overline{S_{\ssm}}.\label{eq:SQC}
\end{align}
Throughout this work, an overline denotes an average over repeats of the experiment. The first equality in Eq.~\eqref{eq:SQC} follows from the fact that $\overline{\rho^{\ssS}_{\ssm}w_{\ssm}}=\overline{\rho_{\ssm}w_{\ssm}}$ for any (matrix-valued) weights $w_{\ssm}$ that can depend on $\ssm$ but that are independent from $V_A,V_B,m_A$, and $m_B$. Using Eq.~\eqref{eq:SQC} we can bound the measurement-averaged von Neumann entropy. Note that since $\rho_{\ssm}$ is a state of just two qubits, the variance of $S^{\ssS\ssC}_{\ssm}$ is in general of order unity, so the average converges rapidly.  

The states $\rho_{\ssm}$ are mixed because of noise in the system, so the von Neumann entropy does not provide us with an entanglement measure. Instead, we probe mixed-state entanglement using the negativity \cite{vidal2003computable}, which vanishes for separable states. The entanglement negativity between $A$ and $B$ in $\rho_{\ssm}$ can be expressed as $N_{\ssm} = -\text{Tr}[ \Pi(\rho^{\textsf{T}_A}_{\ssm}) \rho^{\textsf{T}_A}_{\ssm}]$, where $\rho^{\textsf{T}_A}_{\ssm}$ is the partial transpose of $\rho_{\ssm}$ with respect to degrees of freedom in $A$, and $\Pi(X)$ is the projector onto the span of the eigenstates of the matrix $X$ having negative eigenvalues. We can lower bound the measurement-averaged negativity using \cite{garratt2024probing}
\begin{align}
	\overline{N^{\ssS \ssC}_{\ssm}} = \overline{N^{\ssQ \ssC}_{\ssm}} \leq \overline{N_{\ssm}}, \label{eq:NQC}
\end{align}
where $N^{\ssS \ssC}_{\ssm} = -\text{Tr}[ (\rho^{\ssS}_{\ssm})^{\textsf{T}_A} \Pi((\rho^{\ssC}_{\ssm})^{\textsf{T}_A})]$, and $N^{\ssQ \ssC}_{\ssm}$ is defined by replacing $\rho^{\ssS}_{\ssm}$ in this expression with $\rho_{\ssm}$. If $\overline{N^{\ssS \ssC}_{\ssm}}$ is greater than zero, our measurements must create entanglement between $A$ and $B$.

The inequalities in Eqs.~\eqref{eq:SQC} and \eqref{eq:NQC} are sensitive to the computational model that we use, and are saturated in the idealized case where we have perfect knowledge $\rho^{\ssC}_{\ssm}=\rho_{\ssm}$ of the post-measurement state. More generally, poor estimates $\rho^{\ssC}_{\ssm}$ for $\rho_{\ssm}$ lead to large gaps between our cross-correlations and the true physical quantities, but nevertheless provide physically meaningful bounds. 

\textbf{Machine learning. ---}
Although the above cross-correlations provide bounds on intrinsic properties of post-measurement states, they require computational models $\rho^{\ssC}_{\ssm}$ for the post-measurement states. In order to determine whether the effects of measurements on entanglement are visible in experimental data alone, we ask whether a computational model trained on this data can learn to generate predictions $\rho^{\ssC}_{\ssm}$ for $\rho_{\ssm}$ such that our lower bound on MIE in Eq.~\eqref{eq:NQC} is positive.

In training, the observed two-qubit state of the probe qubits $\ket{\psi_{\ssm}}=\ket{\psi_{\ssm,A}}\otimes \ket{\psi_{\ssm,B}}$ is combined with the output $\rho^{\ssC}_{\ssm}$ of the model in order to construct a loss function, here the negative log-likelihood $-\log_2 \braket{\psi_{\ssm}|\rho^{\ssC}_{\ssm}|\psi_{\ssm}}$ of observing $(m_A,m_B)$ conditioned on the model predicting $\rho^{\ssC}_{\ssm}$. By averaging this loss function over subsets of observed $\ssm$, and varying model parameters to minimize the result, the model may learn to improve the predictions $\rho^{\ssC}_{\ssm}$. After training, we use the model to generate $\rho^{\ssC}_{\ssm}$ for $\ssm$ outside of the training set. Combining $\rho^{\ssC}_{\ssm}$ with the corresponding $\rho^{\ssS}_{\ssm}$, we use Eq.~\eqref{eq:SQC} to upper bound the entropy and Eq.~\eqref{eq:NQC} to lower bound the negativity.

The generative neural networks that we focus on feature an attention mechanism and are inspired by the BERT language model \cite{devlin-etal-2019-bert}. For the one-dimensional array we also construct models $\ssm \mapsto \rho^{\ssC}_{\ssm}$ via variational optimization of tensor networks (Methods).

Crucially, here all training is unsupervised. Previous works have demonstrated how the supervised training of neural networks can be used to study measurement-induced phenomena \cite{dehghani2023neural,agrawal2024observing,kim2025learning}. However, supervised learning requires access to properties of the system that are not necessarily available in experimental observations.

We will contrast results obtained using machine learning models with results obtained using models for the unitary gates used to prepare the state. Neglecting errors in state preparation, these gate-based models generate pure predictions $\hat \rho_{\ssm}$ for post-measurement states of probe qubits. We then depolarize these states in classical post-processing, using states $\rho^{\ssC}_{\ssm} = (1-\epsilon)\hat \rho_{\ssm} + (\epsilon/4)\mathds{1}$ with $\epsilon=0.3$ \footnote{The value $\epsilon=0.3$ is chosen as we find that this improves our bounds for both one- and two-dimensional arrays across wide ranges of parameters.} to construct bounds on properties of post-measurement states.

\begin{figure*}
\includegraphics[width=\textwidth]{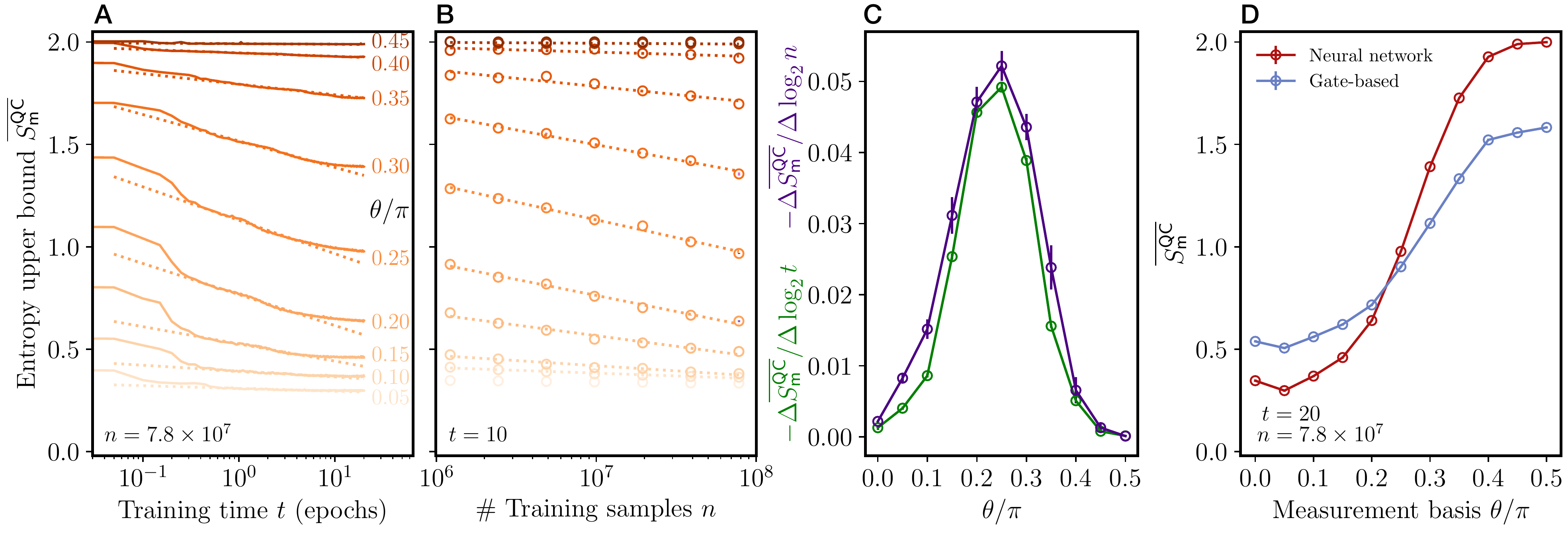}
\caption{Learning post-measurement states of two-dimensional cluster states of $6 \times 6$ qubit arrays. Probe qubits are separated by distance $d=4$ along one edge of the square as illustrated in Fig.~\ref{fig:diagram}\textsf{C}. \textsf{A}: Decrease in $\overline{S^{\ssQ\ssC}_{\ssm}}$ with increasing training time $t$. Different colors correspond to different measurement bases, parameterized by $\theta$, with $\theta/\pi = 0.05, 0.1, \cdots, 0.45$ increasing from bottom to top. During each of the $t=20$ epochs the NN accesses all $n_{\rm{max}} = 7.8 \times 10^7$ training samples. \textsf{B}: Decrease in $\overline{S^{\ssQ\ssC}_{\ssm}}$, at $t=10$ epochs, as the number of training samples $n$ accessed during each epoch is increased (note that in each epoch the NN accesses the same set of samples). The dotted lines show logarithmic decrease of $\overline{S^{\ssQ\ssC}_{\ssm}}$ with the number of samples. \textsf{C}: Sensitivity of $\overline{S^{\ssQ\ssC}_{\ssm}}$ to changes in $t$ (green) and $n$ (purple). Here the gradient $\Delta\overline{S^{\ssQ\ssC}_{\ssm}}/\Delta \log_2 t$ is determined via least-squares fit to data in \textsf{A} from $t=0.5$ to $t=5$ (dotted lines), and $\Delta\overline{S^{\ssQ\ssC}_{\ssm}}/\Delta \log_2 t$ is obtained from a fit to the full range of $n$ in \textsf{B}. In Ref.~\cite{SI} we show that, for each $\theta$, $\overline{S^{\ssQ\ssC}_{\ssm}}$ collapses onto a function of the product $nt$. \textsf{D}: $\overline{S^{\ssQ \ssC}_{\ssm}}$ computed from (red) a NN trained for $t=20$ epochs on $n=n_{\rm{max}}$ samples and (blue) from cross-correlations between a noiseless simulation of the post-measurement state and experimental data.}
\label{fig:vN2d}
\end{figure*}

\textbf{One dimension. ---}
In Fig.~\ref{fig:chain} we demonstrate MIE for the one-dimensional cluster state using a variety of approaches. First, we use gate-based models $\rho^{\ssC}_{\ssm}$ to construct lower bounds $\overline{N^{\ssQ \ssC}_{\ssm}}$ on the average negativity, finding $\overline{N_{\ssm}} > 0$ for the longest chains studied (consisting of $L=34$ qubits). In the inset we determine an upper bound $\overline{S^{\ssQ \ssC}_{\ssm}}$ on the measurement-averaged entropy $\overline{S_{\ssm}}$. Note that the decrease of $\overline{N^{\ssQ \ssC}_{\ssm}}$ with $L$ coincides with an increase in $\overline{S^{\ssQ \ssC}_{\ssm}}$; this behavior is consistent with environmental decoherence being more significant in larger systems.

Although the gate-based scheme detects entanglement, it requires advance knowledge of the quantum state. Our unsupervised learning approach does not suffer from this problem. Lower bounds on negativity based on learning from data, also shown in Fig.~\ref{fig:chain}, demonstrate that the entanglement generated is visible in the experimental data alone. There we show results obtained using attention-based NNs, as well as variationally optimized tensor network models. The training data consists of $8 \times 10^5$ sets of outcomes $\ssm$ and corresponding shadows $\rho^{\ssS}_{\ssm}$ (far fewer than the $2^{32} \approx 4.3 \times 10^9$ different possible $\ssm$), and lower bounds on negativity are constructed by cross-correlating predictions of the trained model with $10^5$ `test' repeats, again each consisting of $\ssm$ and $\rho^{\ssS}_{\ssm}$. Remarkably, the lower bounds on negativity that we obtain using these data-driven models are comparable to those obtained given knowledge of the quantum state. 

In general, long-distance conditional correlations that depend on an extensive number of measurement outcomes are expected to be hard to learn \cite{hu2025local,kumar2026unlearnable}. Even in our one-dimensional system where the NN only needs to estimate a global parity function, learning via gradient descent is extremely costly \cite{shalevschwartz2017failures}. To mitigate this issue, in Fig.~\ref{fig:chain} we trained NNs for increasing system sizes sequentially, using NN parameters obtained from training at size $L$ as the initial parameters for the NN at size $L+2$.

\begin{figure}
\includegraphics[width=0.47\textwidth]{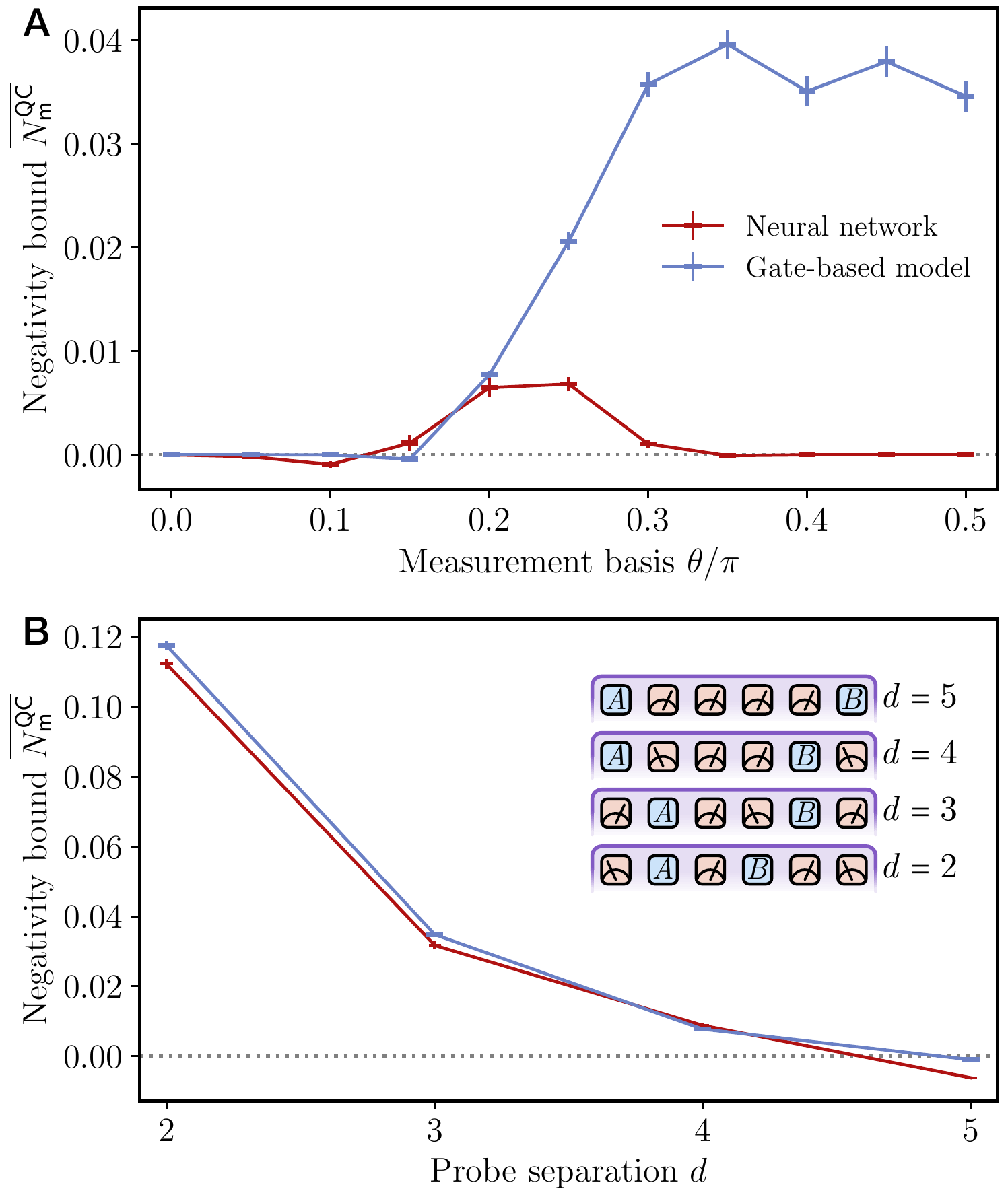}
\caption{Measurement-induced mixed-state entanglement in two-dimensional cluster states of $6 \times 6$ qubit arrays. Both panels show the lower bound $\overline{N^{\ssQ\ssC}_{\ssm}}$ on measurement-averaged entanglement negativity $\overline{N_{\ssm}}$. Lower bounds are obtained by cross-correlating experimental data with (red) predictions of the attention-based neural network after $t=20$ epochs of training and (blue) a gate-based simulation of the effects of measurements on the cluster state. \textsf{A}: Probe qubits at separation $d=4$, for various measurement bases $\theta$. \textsf{B}: Probe qubit at various separations $d$, and with measurement basis $\theta/\pi = 0.2$. The diagram shows the locations of the probe qubits $A$ and $B$ within the upper row of the $6 \times 6$ qubit array.}
\label{fig:N2d}
\end{figure}

\textbf{Two dimensions. ---}
We now turn to the detection of MIE in a two-dimensional cluster state. Here, as the measurement basis is varied, we find a transition in the ability of a neural network to generalize from data. This transition is related to the measurement-induced phase transition \cite{skinner2019measurement,li2018quantum}, which is known to arise when measuring two-dimensional tensor network states \cite{napp2022efficient}. First we study the process of learning the computational model $\ssm \mapsto \rho^{\ssC}_{\ssm}$ from data, and the results are shown in Fig.~\ref{fig:vN2d}. Following this we use the model to detect negativity, and the results are shown in Fig.~\ref{fig:N2d}. 

In Figs.~\ref{fig:vN2d}\textsf{A} and \textsf{B} we study the effects of increasing the number of epochs $t$ and samples $n$ used in training, respectively.  Since $\overline{S^{\ssQ\ssC}_{\ssm}}=\overline{S_{\ssm}}+\overline{D^{\mathrm{KL}}_{\ssm}}$ any decrease in $\overline{S^{\ssQ\ssC}_{\ssm}}$ is a decease in $\overline{D^{\mathrm{KL}}_{\ssm}}$ and hence an improvement in the model $\ssm \mapsto \rho^{\ssC}_{\ssm}$. After training for $t$ epochs on $n$ samples, $\overline{S^{\ssQ\ssC}_{\ssm}}$ is computed from cross-correlations between the model and the shadows $\rho^{\ssS}_{\ssm}$ extracted from a separate $10^6$ `test' repeats of the experiment. Crucially, the maximum number $n_{\rm{max}} = 7.8 \times 10^7$ of training samples is also orders of magnitude smaller than the total number $2^{34} \approx 1.7 \times 10^{10}$ of possible outcomes $\ssm$ in the $6 \times 6$ array, so we are in a regime where postselection is not feasible.

In Fig.~\ref{fig:vN2d}\textsf{A} we show results from $t=0.05$ to $t=20$ epochs of training time for $n = n_{\rm{max}}$. For small $\theta$, already after $0.05$ epochs, $\overline{S^{\ssQ\ssC}_{\ssm}}$ is relatively small, suggesting that the estimates $\rho^{\ssC}_{\ssm}$ are good approximations to the true post-measurement density matrices $\rho_{\ssm}$. On the other hand, for large $\theta$, even after $t=20$ epochs, $\overline{S^{\ssQ\ssC}_{\ssm}}$ is close to two bits, which is the value expected when $\rho^{\ssC}_{\ssm}$ is maximally mixed. This is because, at large $\theta$, the state $\rho_{\ssm}$ depends sensitively on an extensive number of measurement outcomes and hence the NN is not able to approximate the relation between $\ssm$ and $\rho^{\ssC}_{\ssm}$. At intermediate $\theta$ we observe a gradual decay of $\overline{S^{\ssQ\ssC}_{\ssm}}$ over many epochs. This suggests that post-measurement states are highly structured, but that the NN is still able to learn. 

In Fig.~\ref{fig:vN2d}\textsf{B}, instead fixing $t=10$ and increasing $n$, we find an unanticipated logarithmic decrease of $\overline{S^{\ssQ\ssC}_{\ssm}}$ with $n$. There are also remarkable similarities with Fig.~\ref{fig:vN2d}\textsf{A}: the effect of increasing either $n$ or $t$ appears to have a much more significant impact at intermediate $\theta$. This variation is a signature of a transition in the ability of a classical computational model to learn and to generalize from data. 

We explore this signature in more detail in Fig.~\ref{fig:vN2d}\textsf{C}. There we show that the rates of changes of the quantum KL divergence with $\log_2 n$ and with $\log_2 t$ exhibit peaks with similar profiles and magnitudes. Moreover, in Ref.~\cite{SI} we show that $\overline{S^{\ssQ \ssC}_{\ssm}}$, extracted for various $t$ and $n$, approximately collapses onto a ($\theta$-dependent) function of $nt$, which is proportional to the total computational time used in training. This collapse can be reconciled with the differences between Figs.~\ref{fig:vN2d}\textsf{A} and \textsf{B} due to the fact that, for fixed $n$, the behavior of the upper bound $\overline{S^{\ssQ \ssC}_{\ssm}}$ becomes complicated at large $t$ due to overfitting. 

This result has important implications for the resources required to detect a learnability transition in large systems. Based on Figs.~\ref{fig:vN2d}\textsf{B} and \textsf{C}, and the collapse in Ref.~\cite{SI}, let us assume that $\overline{S^{\ssQ \ssC}_{\ssm}} \approx a-b\log_2 (nt)$ during the late stages of training. If we also assume that $\overline{S^{\ssQ \ssC}_{\ssm}}$ converges to the final value $\overline{S_{\ssm}}$ (in absence of noise this is zero) for $n$ and $t$ that are no larger than exponential in (a power of) the number $L^2$ of qubits, then $b$ cannot decrease with $L$ faster than $L^{-2}$. The logarithmic rate of change $b = -\Delta\overline{S^{\ssQ \ssC}_{\ssm}}/\Delta \log_2 n$ can therefore be determined, to fixed relative error, from a number of repeats of the experiment that is polynomial in the system size. As we have shown in Fig.~\ref{fig:vN2d}\textsf{C}, this object exhibits a distinct peak at the transition.

In Fig.~\ref{fig:vN2d}\textsf{D} we compare the entropy upper bounds  $\overline{S^{\ssQ\ssC}_{\ssm}}$ obtained using generative NN models with the upper bounds obtained using gate-based models. Interestingly, the NN gives a tighter bound at small $\theta$; one possibility is that it is able to learn gate calibration errors and correlations in the noise that are absent in our gate-based model. At large $\theta$ the NN gives $\overline{S^{\ssQ\ssC}_{\ssm}} \approx 2$ for reasons discussed above, but the gate-based model gives a significantly smaller value. This is only possible if the NN has failed to learn structure that is in fact present in the post-measurement states. 

Having characterized the learning process, we now use the models trained for $t=20$ epochs (on $n=n_{\rm{max}}$ samples) to detect negativity in the experimental post-measurement states. The results of our experiments on MIE in two-dimensional cluster states are shown in Fig.~\ref{fig:N2d}. In Fig.~\ref{fig:N2d}\textsf{A} we show results for various bases $\theta$ and for probe qubits separated by $d=4$ [see Fig.~\ref{fig:diagram}B]. Using the machine learning model we find a peak, with $\overline{N^{\ssQ\ssC}_{\ssm}}>0$ and hence $\overline{N_{\ssm}}>0$, at intermediate $\theta$. This is one of our central results: a key signature of a measurement-induced phase transition is visible in experimental data alone, and its observation requires neither postselection nor advance information on the structure of the unmeasured state. Previous experiments have successfully detected signatures of such transitions in statistical properties of observed measurement outcomes \cite{noel2022measurement,google2023measurement,kamakari2025experimental}, but it unclear whether they have detected entanglement. Moreover, all of these experiments required advance knowledge of the protocol used to prepare the quantum state, or otherwise postselection from an exponentially large number of trial runs \cite{koh2023measurement}. In Fig.~\ref{fig:N2d}\textsf{B} we then show the effect of varying $d$ at fixed $\theta/\pi=0.2$: although there is MIE between probe qubits for $d\leq 4$, for $d=5$ we do not detect MIE.

Note that the peak in $\overline{N^{\ssQ\ssC}_{\ssm}}$ obtained using the NN occurs over approximately the same window of $\theta$ as the peaks in Fig.~\ref{fig:vN2d}\textsf{C}. This behavior is in line with the theoretical expectation that MIE first emerges at the threshold between learnable and unlearnable post-measurement states. Comparing with the lower bound $\overline{N^{\ssQ\ssC}_{\ssm}}$ computed using a gate-based model, we see that MIE survives at larger values of $\theta$, extending all the way up to $\theta/\pi=0.5$, but this structure is invisible to the NN. At large $\theta$ the function $\ssm \mapsto \rho_{\ssm}$ is sufficiently complex that the NN is not able to generalize from previous observations and recognize MIE. 

The connection between the entanglement and learning transitions can be understood in part from studies of the conditional mutual information (CMI) of classical distributions \cite{yang2024classical,kumar2026unlearnable,hu2025local}. MIE between the probes necessitates CMI in the classical probability distribution over sets of outcomes that we sample when measuring all of the qubits. At small $\theta$ the CMI decays exponentially over a finite Markov length, while at the transition the Markov length diverges. The success of our unsupervised learning scheme at small $\theta$ is consistent with the idea that a finite Markov length implies an efficient NN representation \cite{yang2024classical} and, in some cases, efficient learning \cite{hu2025local,kumar2026unlearnable}. At large $\theta$, on the other hand, learning is not scalable. An open question is whether MIE can be scalably detected at the entanglement transition.

\textbf{Outlook. ---}
Our experiments demonstrate how we can observe the effects of large numbers of measurements on a quantum system. A key initial step is to learn how to infer the effects of measurements on unmeasured quantum degrees of freedom. Once an approximate model for the system is generated from training data, cross-correlations between the model and an independent dataset can be used to construct bounds on properties of measured quantum states. These protocols highlight a deep connection between our ability to learn how to model a quantum system, and our ability to observe the effects of measurements.

Because our method only requires us to repeatedly prepare the initial quantum state but does not require us to have an accurate model for this state (in contrast with Refs.~\cite{noel2022measurement,google2023measurement,agrawal2024observing,kamakari2025experimental}) it allows for the study of measurement-induced collective phenomena in general settings. It is important to note also that, although we have here focused on entanglement, cross-correlations can be used to infer the effects of measurements on more standard observables \cite{garratt2024probing,mcginley2024postselection}. For example, our method allows for the study of measurement-induced phenomena in ultracold atomic and molecular systems, where quantum gas microscopes provide high-resolution images of particle locations \cite{gross2021quantum}, and where we do not have a precise description of quantum state preparation. 

An immediate application is to quantum control, in particular quantum error correction \cite{shor1995scheme,terhal2015quantum}. Previous experiments on quantum error correction with stabilizer codes \cite{krinner2022realizing,sivak2023real,google2023suppressing,bluvstein2024logical}, as well as on the preparation of topologically ordered quantum states via measurement \cite{iqbal2024topological,iqbal2024non,xu2024non}, have focused on highly controlled settings where the relation between outcomes and post-measurement states corresponds to a known classical algorithm. Using the tools developed here future works can move beyond these settings and explore the behavior of quantum memories when they are far from commuting stabilizer limits, and also to develop error correction schemes that are tailored to specific quantum algorithms.

\bigbreak

\textbf{Acknowledgements. ---} This work was supported by NSF Grant No. DMR-2238360 (WH and YZY), the Gordon and Betty Moore Foundation (SJG), the NSF QLCI program through Grant No. OMA-2016245 (EA), and a Simons Investigator Award (EA). The data presented in Figs.~\ref{fig:vN2d} and \ref{fig:N2d} were taken remotely on a 105-qubit Willow processor \cite{acharya2024quantum}, with access provided via Google's Quantum Engine. Calibration and support were provided by the Quantum Hardware Residency Program. We thank the Google Quantum AI team for providing the quantum systems and support that enabled these results. We thank Michael Broughton for his comments on the draft. The views expressed in this work are solely those of the authors and do not reflect the policy of Google or the Google Quantum AI team.

\bigbreak

\textbf{Data and code availability. ---} Python scripts for quantum state preparation and experimental data collection, based on the Cirq framework, are available at Ref.~\cite{cirqcode}. Additional scripts used for machine learning and gate-based models are available at Ref.~\cite{hou2025mlmipt}.

%

\newpage

\section*{Methods}

Here we provide additional details on our experimental and numerical methods. First, in Sec.~\ref{sec:circuit} we describe experimental state preparation and error detection. In Sec.~\ref{sec:MLsetup} we outline the machine learning methods used to extract models for post-measurement states from data. In Sec.~\ref{sec:noiseless} we describe the gate-based models for post-measurement states. 

\section{Experimental protocols}\label{sec:circuit}

Here we describe the experimental preparation of initial states, the measurements used to restructure these states, and our strategy for detecting measurement errors. Section~\ref{subsec:detect} describes error detection, and Sections~\ref{subsec:tel} and ~\ref{subsec:cluster} describe the protocols used for the one- and two-dimensional arrays, respectively. The scripts used for our experiments are available at Ref.~\cite{cirqcode}.

Although we are interested in the effects of measurements on the states of probe qubits $A$ and $B$, in our experiments we perform all measurements simultaneously. These include the measurements used to prepare the states $\rho_{\ssm}$ on $A$ and $B$, as well as the measurements used to characterize $\rho_{\ssm}$. Our experiments all involve the following steps, and start with all qubits initialized in the state $\ket{0}$:
\begin{enumerate}[(i)]
\item Apply two-qubit unitary operation to prepare an entangled many-qubit state.
\item Apply single-qubit unitary operations that determine the measurement basis. 
\item Entangle system qubits with error detection qubits.
\item Measure all qubits in the computational basis.
\end{enumerate}
The random bases in which we measure the probe qubits are defined by the unitary operations $V_A$ and $V_B$ applied in step (ii); these operations are randomly and independently sampled from the set $\{ \mathds{1},e^{i\frac{\pi}{4}X},e^{i\frac{\pi}{4}Y}\}$.

\subsection{Error detection}\label{subsec:detect}
Errors in the outcomes of measurements have a significant impact on our results, so we implement a simple error detection scheme. In step (iii) we introduce error-detection qubits initialized in state $\ket{0}$, and apply $\mathrm{CNOT}$ gates each having a system qubit as control and a distinct error-detection qubit as target. In practice this is implemented as $\mathrm{CNOT} = (\mathds{1} \otimes H) \cdot \mathrm{CZ} \cdot (\mathds{1} \otimes H)$. Note that not all system qubits are associated with error-detection qubits: in one-dimensional arrays we only introduce error-detection qubits for the probes $A$ and $B$, and in two-dimensional arrays we introduce error-detection qubits around the perimeter of the square array. In the two-dimensional arrays this means that qubits at an edge have one error-detection qubit, whereas qubits at corners have two error-detection qubits. 

When we measure all qubits in step (iv), we can then identify measurement errors by comparing outcomes on error-detection qubits and their associated system qubits. In particular, if the outcome observed on an error-detection qubit does not match its corresponding system qubit, there must have been an error. In that case we discard this run of the experiment. This corresponds to postselecting on error-free repeats of experiment. 

\subsection{One-dimensional array}\label{subsec:tel}

\begin{figure}[htbp]
\begin{center}
\includegraphics[width=0.47\textwidth]{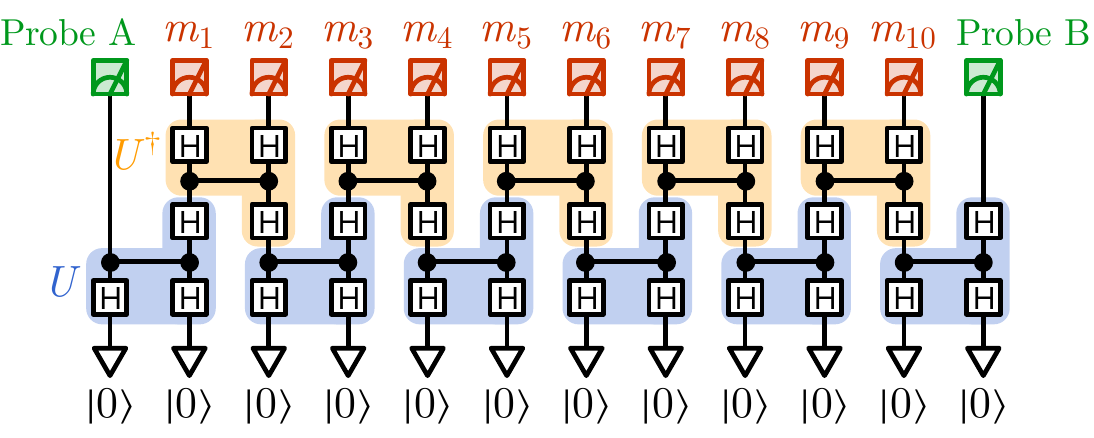}
\caption{Experimental realization of the quantum circuit that creates one-dimensional cluster states, and the measurements (red) in the bulk used to generate entanglement between probe qubits at the ends of the chain (green). The green measurements of the probe qubits are performed in random bases, and the results are used to construct classical shadows.}
\label{fig:tel_circuit}
\end{center}
\end{figure}

Experiments on one-dimensional arrays were performed using a Google Sycamore chip, and a detailed analysis of the hardware can be found in Ref.~\cite{arute2019quantum}. For our experiments we choose a one-dimensional array of qubits which `snakes' through the two-dimensional array of qubits, and we apply a depth-$2$ circuit to this chain to prepare a one-dimensional cluster state. The circuit for state preparation in step (ii) is illustrated in the yellow and blue boxes in Fig.~\ref{fig:tel_circuit}. The two-qubit gate described in the main text was experimentally implemented as $U = (H \otimes H) \cdot \mathrm{CZ} \cdot (\mathds{1} \otimes H)$, as indicated in the blue regions. These gates, which act on pairs of qubits $(2j-1,2j)$ for $j=1,2,\cdots$, are followed by gates $U^{\dagger}$ which act on pairs of qubits $(2j,2j+1)$.  

In the absence of noise, measurements of the non-probe qubits would prepare the two probe qubits into one of four standard EPR pairs, which we denote by $\ket{AB}$. Exactly which state is determined by the parity of the measurement results $m_2,\ldots,m_{L-1}$. With outcomes $m_i=\pm 1$ the four EPR pairs are indexed by $x = \prod_{i \in \text{even}} m_i$ and $y = \prod_{i \in \text{even}} m_i$; explicitly,
\begin{align}
    \ket{AB} = \frac{1}{\sqrt{2}}\Big( \ket{0}\otimes \ket{(x-1)/2} + y \ket{1}\otimes \ket{(x+1)/2}\Big),\notag
\end{align}
where addition in the state is defined modulo $2$, i.e. $\ket{-1}=\ket{1}$.

\subsection{Two-dimensional array}\label{subsec:cluster}

\begin{figure}[htbp]
\begin{center}
\includegraphics[width=0.35\textwidth]{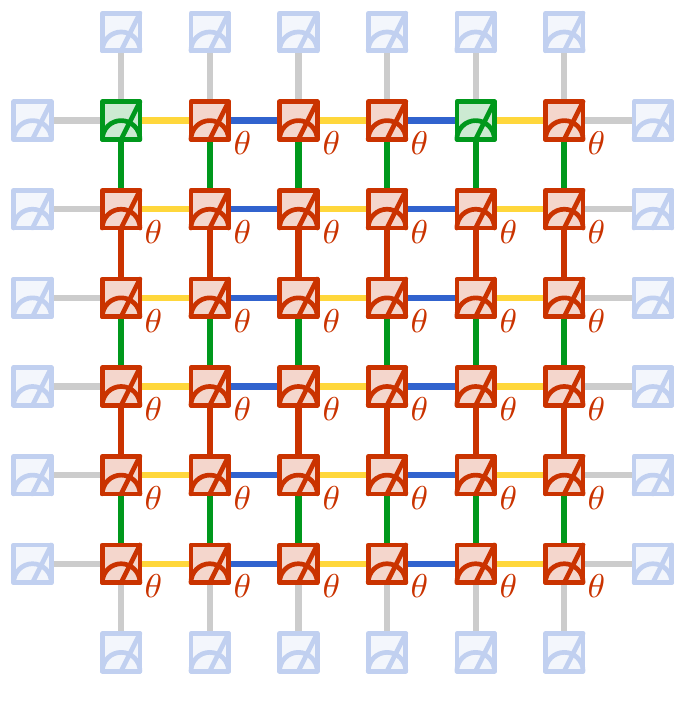}
\caption{Experiments on two-dimensional cluster states. The cluster state (on red and green qubits) is prepared by applying Hadamard gates followed by a sequence of two-qubit gates $\exp[i (\pi/4)\sum_{\langle j,k \rangle} Z_j Z_k]$ between pairs of qubits $(j,k)$ connected by lines marked (I) yellow, (II) blue, (III) green, and then (IV) red. Following this we apply single qubit gates to change the measurement bases, and then $\mathrm{CNOT}$ gates having system qubits at the edges of the square as controls and error-detection qubits (faded blue) as targets. Finally we measure all qubits in the computational basis.}
\label{fig:shallow_lattice}
\end{center}
\end{figure}

Experiments on one-dimensional arrays were performed using a Google Willow chip, discussed in detail in Ref.~\cite{acharya2024quantum}. This section describes our experiments on two-dimensional arrays. We denote by $L=6$ the system linear dimension, with $L^2=36$ the total number of system qubits. After initializing these qubits in state $|0\rangle^{\otimes L^2}$, the step (ii) involves applying the following sets of gates (in the order listed)
\begin{enumerate}
    \item Hadamard: $\bigotimes_j H_j$
    \item Nearest-neighbor $ZZ$: $\exp[i (\pi/4)\sum_{\langle j,k \rangle} Z_j Z_k]$
    \item Single-qubit $Z$: $\bigotimes_j \exp[i(\phi/2) Z_j]$.
    \item Single-qubit $Y$: $\bigotimes_j \exp[i(\theta/2) Y_j]$
\end{enumerate}
The nearest-neighbor gate in step 2,  
$\exp[i (\pi/4)\sum_{\langle j,k \rangle} Z_j Z_k]$, 
is implemented by first applying a controlled-$Z$ ($\mathrm{CZ}$) gate between each pair of qubits, followed by local $Z^{-\frac{1}{2}}$ operations on both qubits, i.e., 
$Z_j^{-1/2}Z_k^{-1/2}\mathrm{CZ}_{jk}$. All nearest-neighbor two-qubit gates were applied in the following sequence: (I) odd horizontal links (II) even horizontal links (III) odd vertical links, and finally (IV) even vertical links. This sequence of two-qubit gates is indicated by the different colored lines in Fig.~\ref{fig:shallow_lattice}. Additionally, $4L$ error-detection qubits surround the $L \times L$ array.

\section{Machine learning models}\label{sec:MLsetup}
In this section we discuss the use of unsupervised learning (in particular, self-supervised learning) to generate estimates $\rho^{\ssC}$ for post-measurement density matrices $\rho_{\ssm}$. In Sec.~\ref{subsec:TranNN}, we describe the scheme used for the results presented in the main text, where we use an attention-based neural network (NN) to generate models from observed sets of outcomes $\ssm_r$ and shadows $\rho^{\ssS}_{r}$. In Sec.~\ref{subsec:mpo} then describe the scheme based on variational optimization of tensor networks that we use to characterize post-measurement states in one-dimensional qubit arrays. 

First, however, we provide a high-level description of our approach. We denote by $r=1,\cdots,R_{\text{tr}}$ a set of repeats of the experiment that is used for training, with corresponding sets of outcomes $\ssm_r$, and by $\lambda$ the set of internal parameters of the neural network. The set of outcomes $\ssm_r$ observed in run $r$ (which does not include any information about measurements on probe qubits) is provided as a prompt, and the NN produces a valid density matrix $\rho^{\ssC}_{\ssm_r}(\lambda)$. From the single-qubit unitary operations $V_{A,r}$ and $V_{B,r}$ that we apply to the probe qubits, and the measurement outcomes $m_{A,r}$ and $m_{B,r}$ observed on these qubits, we also construct the pure states $\ket{\psi_{A,r}}=V_{A,r}^{\dag}\ket{m_{A,r}}$ and $\ket{\psi_{B,r}}=V_{B,r}^{\dag}\ket{m_{B,r}}$, with $\ket{\psi_r}=\ket{\psi_{A,r}}\otimes \ket{\psi_{B,r}}$. The pure states $\ket{\psi_r}$ and the density matrices $\rho^{\ssC}_{\ssm_r}$ are used to compute a loss function, here the negative log-likelihood (NLL). In run $r$, the NLL is
\begin{align}
    \text{NLL}_r(\lambda) = - \log \braket{\phi_r|\rho^{\ssC}_{\ssm_r}(\lambda)|\phi_r}.
\end{align}
In practice we average $\text{NLL}_r(\lambda)$ over a minibatch (i.e. over a small subset of the $R_{\text{tr}}$ repeats used for training) and, after evaluating derivatives of $\text{NLL}_r(\lambda)$ with respect to the parameters $\lambda$, we update $\lambda$ via stochastic gradient descent. The final set of parameters in this scheme, which denote $\lambda^*$, defines the model that is then used to construct bounds on measurement-induced entanglement and the von Neumann entropy, i.e. $\rho^{\ssC}_{\ssm} = \rho^{\ssC}_{\ssm}(\lambda^*)$.

\begin{figure}[htbp]
\centering
\includegraphics[width=0.47\textwidth]{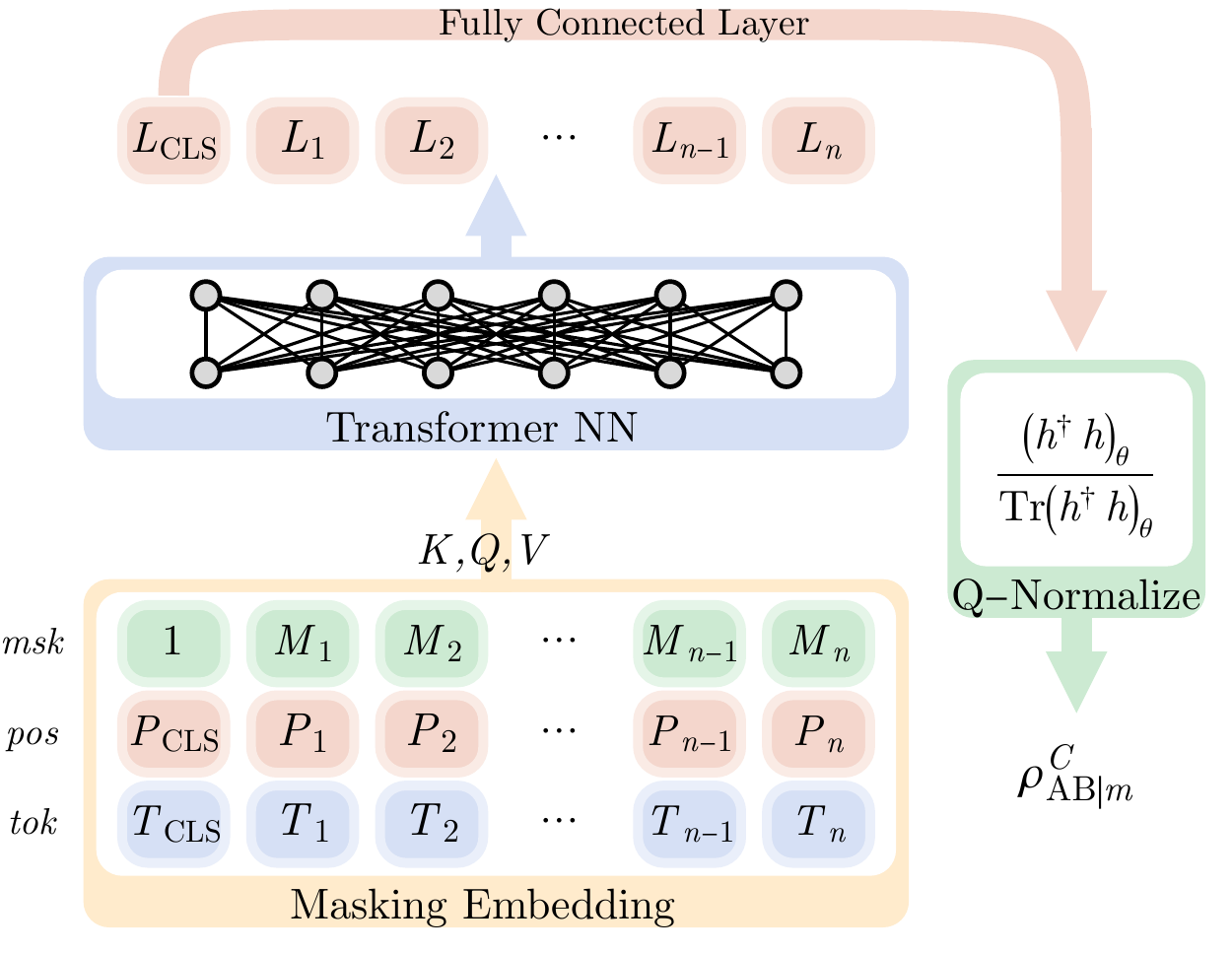}
\caption{Transformer NN for predicting the probe qubit state from preparation measurements. The input is tokenized and positionally encoded, then processed by the Transformer. The CLS token is mapped to a $4 \times 4$ complex matrix $h$, converted into a valid density matrix $\rho^{C}_{\text{AB}|m}$ via a quantum normalization layer.}
\label{fig:TranNN}
\end{figure}

\subsection{Transformer Neural Network}\label{subsec:TranNN}

The attention-based neural networks that we use to learn computational models $\ssm \mapsto \rho^{\ssC}_{\ssm}$ are inspired by the BERT model used in natural language processing \cite{devlin-etal-2019-bert}. Figure~\ref{fig:TranNN} illustrates the full architecture. The source code and trained models described here are available at Ref.~\cite{hou2025mlmipt}, and the transformer neural network architecture is implemented using the Hugging Face Transformers library \cite{transformers2024}.

The set of binary measurement outcomes $\ssm$, the positions of these measurements in the qubit arrays, and a mask degree of freedom for each measurement (see below) are first embedded into one-dimensional arrays. Each of these three pieces of information is associated with a different subset of elements of this array, indicated by `tok', `pos' and `msk' in Fig.~\ref{fig:TranNN}, respectively. There is one such array for each measurement (i.e. one for each element of $\ssm_r$, so $M$ in total). An additional array, indicated by CLS (in the standard use of the BERT model this is known as a  `classification' token, although here it is not used for classification) is prepended to the set of $M$ arrays described above. The output of the NN is ultimately written into the CLS token.

\begin{figure}[htbp]
\centering
\includegraphics[width=0.47\textwidth]{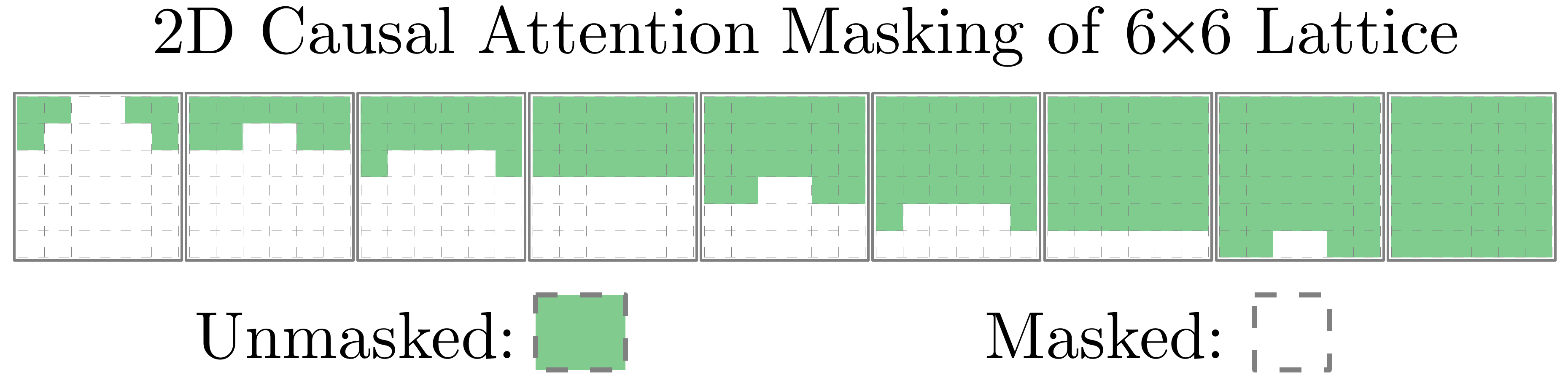}
\caption{Examples of 2D causal masking patterns used during training. In each iteration, the model predicts the probe state using only accessible qubits, starting from the probe site (top opposite corner). This progressive masking enforces locality and regularizes training.}
\label{fig:2Dmask}
\end{figure}

The full set of the $M+1$ one-dimensional arrays is then passed into the NN. During training of the models used in Figs.~\ref{fig:vN2d}\textsf{A} and \textsf{C}, as well as Fig.~\ref{fig:N2d}, we employ a 2D causal attention masking scheme that mimics autoregressive modeling in 1D language tasks. The masking propagates outward from the probe qubit locations, as shown in Fig.~\ref{fig:2Dmask}, limiting each input's `receptive field' based on locality. The idea is to let the NN first attend to the most relevant (nearby) measurements, and then to progressively improve its predictions by exposing more distant measurement outcomes. Note that, after training (i.e. at inference time), masking is disabled and the full measurement sequence is attended to. In Fig.~\ref{fig:vN2d}\textsf{B} we do not use causal attention masking, and this is the reason for the slight differences compared with $t=10$ data in Fig.~\ref{fig:vN2d}\textsf{A}, which are visible for small $\theta$.

After processing through the Transformer layers, the final CLS token representation is passed through a fully connected layer to produce a 32-dimensional output vector. This vector is reshaped into a complex $4 \times 4$ matrix $h_{\ssm}(\lambda)$ depending on the internal parameters $\lambda$ of the NN, and then converted into a valid quantum density matrix of the two probe qubits:
\begin{equation}
\rho^{\ssC}_{\ssm}(\lambda) = \frac{h_{\ssm}^\dagger(\lambda) h_{\ssm}(\lambda)}{\text{Tr}(h_{\ssm}^\dagger(\lambda) h_{\ssm}(\lambda))}.
\end{equation}
During training, $\rho^{\ssC}_{\ssm}(\lambda)$ is then used to compute the NLL, and the parameters $\lambda$ are updated using the Adam optimizer.

\subsection{Matrix-product states}\label{subsec:mpo}

Here we discuss the variational tensor network method used in Fig.~\ref{fig:chain}. This architecture, known as the Born machine \cite{PhysRevX.8.031012}, is based on matrix-product states and so is particularly well-suited to one-dimensional systems.

Our general parameterization and optimization procedures follow Ref.~\cite{2022MLS&T...3a5020G}.

\begin{figure}[htbp]
\centering
\includegraphics[width=0.47\textwidth]{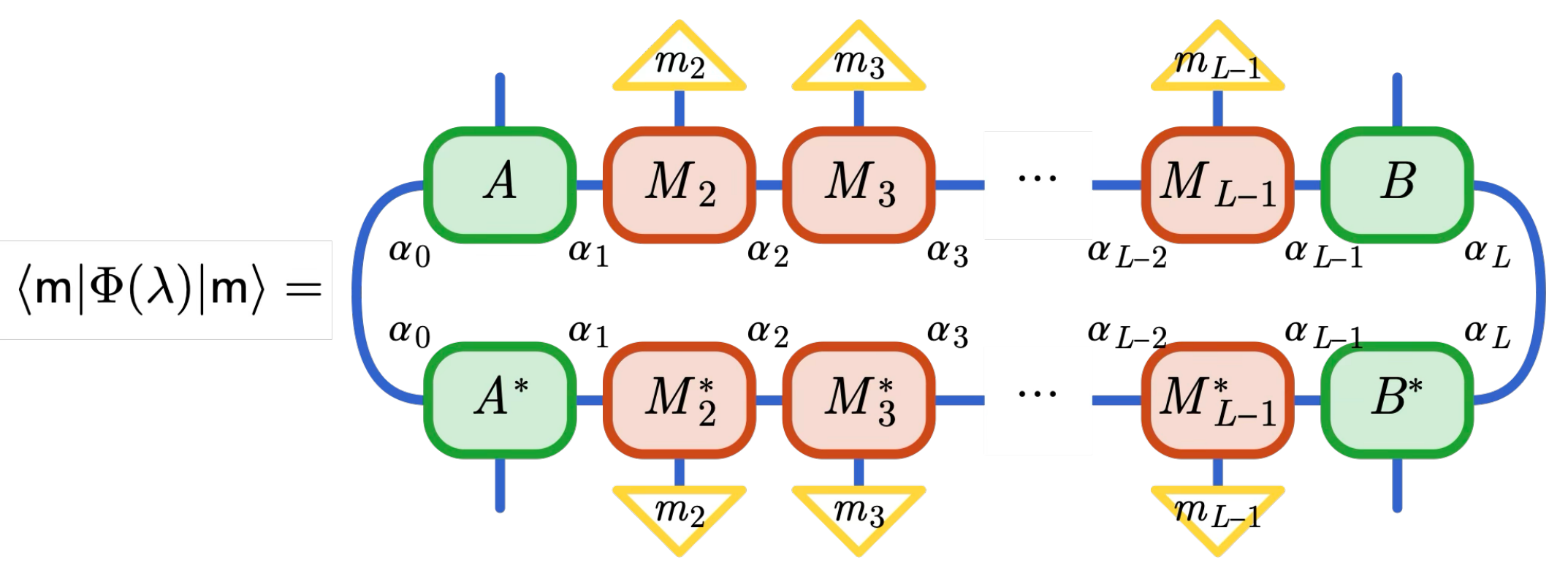}
\caption{Tensor network for constructing $\braket{\ssm|\Phi(\lambda)|\ssm}$. Red tensors $M_i$ can be conditioned on outcomes $m_i$, while $A$ and $B$ are boundary tensors.}
\label{fig:mpo_tn}
\end{figure}

For the one-dimensional array of $L$ qubits, the idea is to construct a computational model $\ssm \to \rho^{\ssC}_{\ssm}$ from a $L$-qubit matrix product operator (MPO) $\Phi(\lambda)$. The set $\lambda$ of internal parameters corresponds to the set of components of the tensors (see below) from which the MPO is constructed. Note that the MPO need not closely resemble the full $L$-qubit state prepared in experiment. It should be viewed only as an intermediate step for the generation of $\rho^{\ssC}_{\ssm}$. Denoting by $\ket{\ssm}=\ket{m_2 \cdots m_{L-2}}$ the product state associated with the measurement outcomes on the non-probe qubits, we have
\begin{align}
    \rho^{\ssC}_{\ssm} = \frac{\braket{\ssm|\Phi(\lambda)|\ssm}}{\text{Tr}\braket{\ssm|\Phi(\lambda)|\ssm}}.
\end{align}
The numerator $\braket{\ssm|\Phi(\lambda)|\ssm}$ of this expression is illustrated in Fig.~\ref{fig:mpo_tn}. The product state $\ket{\ssm}$ is represented as the set of yellow triangles. The $\chi \times 2 \times \chi$ tensors $A$, $B$ (green) and $M_i$ (red) define the MPO $\Phi(\lambda)$ through
\begin{align}
\bra{\ssm}&\Phi(\lambda)\ket{\ssm} = \sum_{\alpha_0,...,\alpha_{L}} 
A^{\alpha_0,\alpha_1} B^{\alpha_{L-1},\alpha_{L}} \\ \times &\left[ \prod_{i=2}^{L-1} M_{m_i}^{\alpha_{i}, \alpha_{i+1}} \left(M_{m_i}^{\alpha_{i}, \alpha_{i+1}}\right)^* \right] \left(B^{\alpha_{L-1},\alpha_{L}}\right)^* \left(A^{\alpha_0,\alpha_1}\right)^*.\notag
\end{align}
As indicated above, $\lambda$ is a compressed description of the components of the tensors $A$, $B$ and $M_i$ tensors. During training, we vary these components in order to minimize the NLL described above.

\section{Gate-based models}\label{sec:noiseless}

In the experimental platform studied here the initial quantum state preparation is well characterized, so we can construct `gated-based models' $\rho^{\ssC}_{\ssm}$ for post-measurement states. In the main text we have used these models to construct quantum-classical cross-correlations, and hence to bound measurement-induced entanglement and the von Neumann entropy. The precise details of the quantum channel which prepares the initial state are not readily available in our measurements, so within this scheme we can detect MIE even when it cannot be efficiently detected by learning from the experimental data alone. 

We construct these models as follows. In run $r$ of the experiment, where we observe a set $\ssm_r$ of measurement outcomes, we construct the corresponding projection operators and apply them to a pure estimate for the initial state (i.e. we neglect errors in the gates). This generates a pure post-measurement state $\hat \rho^{\ssC}_{\ssm_r}$ of the probe qubits, and in classical post-processing we `depolarize' this state to generate the mixed state $\rho^{\ssC}_{\ssm_r}=(1-\epsilon)\hat \rho^{\ssC}_{\ssm_r}+(\epsilon/4)\mathds{1}$. For simplicity we choose $\epsilon=0.3$ throughout this work; this choice improves our bounds over a wide range of parameters, but the bounds could be further improved by optimizing over $\epsilon$. Since our focus in this work is on machine learning models, we do not concern ourselves with this optimization here. 

For one-dimensional arrays the models for post-measurement states described above have a very simple form: $\hat \rho^{\ssC}_{\ssm}$ is an EPR pair. For two-dimensional arrays the post-measurement states are more complicated. Given a set of measurement outcomes $\ssm$, our strategy for numerically calculating $\hat \rho^{\ssC}_{\ssm}$ in a two-dimensional array is as follows:
\begin{enumerate}
    \item Initialize and store the post-Hadamard state of two rows of qubits.
    \item Apply $\exp[i(\pi/4)\sum_{\langle j,k \rangle} Z_j Z_k]$ to all nearest-neighbor qubit pairs $(j,k)$ within and across both rows.
    \item Apply $\exp[i(\theta/2) Y_j]\exp[i(\phi/2) Z_j]$ to all qubits $j$ in one of the rows, and measure that row.
    \item Initialize a new row in the post-Hadamard state. 
    \item Repeat steps 2–4 for a total of $L-1$ iterations. 
    \item On the final row, perform measurements on all qubits except the probes $A$ and $B$.
\end{enumerate}
Using this scheme, we can determine the post-measurement state of the two probe qubits while only ever storing a quantum state of $2L$ qubits (rather than the full $L^2$).

\clearpage

\onecolumngrid
\section*{Supplemental information}

This supplemental information is organized as follows. First, in Sec.~\ref{sec:QC} we explain how the data extracted in experiment can be used to constrain measurement-induced entanglement. In Sec.~\ref{sec:1dextra} we show additional experimental results on one-dimensional arrays; these include the extraction of lower bounds on coherent information, and a test of an alternative method for detecting the effects of large numbers of measurements. Additional experimental results on the two-dimensional array are shown in Sec.~\ref{sec:2dextra}. There we provide evidence that distant measurement outcomes affect the states of the probe qubits generated in experiment, and that the distribution of Born probabilities $p_{\ssm}$ develops nontrivial structure for intermediate bases $\theta$. Finally, in Sec.~\ref{sec:noiseless2}, we present numerical simulations of the effects of measurements on two-dimensional cluster states.

\section{Quantum-classical cross-correlations}\label{sec:QC}
Here we review the cross-correlations introduced in Ref.~\cite{garratt2024probing}, providing a self-contained derivation of those which lower bound mixed-state entanglement measures and which upper bound the entropy. In the main text the mixed-state entanglement measure that we have focused on is the negativity, and here we will also discuss the coherent information; in Sec.~\ref{sec:1dextra} we will present experimental measurements of our lower bound on the measurement-averaged coherent information in one-dimensional cluster states.

The setup is as follows: In each repeat $r$ of the experiment (with $r=1,\ldots,R$ and $R$ the total number of repeats) we perform $M$ measurements, finding a set of outcomes $\ssm_r = (m_{r,1},\ldots,m_{r,M})$ with Born probabiliy $p_{\ssm_r}$. In general $p_{\ssm_r}$ is exponentially small in $M$, and we are interested in the case where $M$ scales with the number of qubits $N$. For $R$ growing no faster than polynomially with $N$, with high probability we therefore have $\ssm_r \neq \ssm_{r'}$ for $r \neq r'$ at large $N$. Because of this, in the main text we have simplified our notation by labelling repeats $r$ of the experiment by the observed outcomes $\ssm_r$; here we keep the label $r$ explicit for concreteness. In repeat $r$ of the experiment we create a post-measurement state $\rho_{\ssm_r}$ of the probe qubits $A$ and $B$ that depends only on $\ssm_r$.

In each repeat we apply random unitary operations $V_{A,r}$ and $V_{B,r}$ to the probe qubits $A$ and $B$, with each of $V_{A,r}$ and $V_{B,r}$ drawn independently from the uniform distribution over $\{\mathds{1},e^{i\frac{\pi}{4}X},e^{i\frac{\pi}{4}Y}\}$. After applying these unitary operations we measure the probe qubits in the computational basis (i.e. the basis of eigenstates of Pauli $Z_A$ and $Z_B$), finding outcomes $m_{A,r}$ and $m_{B,r}$, which correspond to eigenstates of the Pauli $Z_A$ and $Z_B$ operators. From these measurements we compute shadows
\begin{align}
    \rho^{\ssS}_{r,A} = 3 V_{A,r}^{\dag}\ket{m_{A,r}}\bra{m_{A,r}}V_{A,r}-\mathds{1}, \quad \rho^{\ssS}_{r,B} = 3 V_{A,r}^{\dag}\ket{m_{A,r}}\bra{m_{A,r}}V_{A,r}-\mathds{1},
\end{align}
while the two-qubit shadow is simply $\rho^{\ssS}_{r} = \rho^{\ssS}_{r,A} \otimes \rho^{\ssS}_{r,B}$. The probability that the shadow is $\rho^{\ssS}_r$ above, conditioned on having observed outcomes $\ssm_r$ (note that this string of outcomes does not include $m_{A,r}$ and $m_{B,r}$) is 
\begin{align}
    p(\rho^S_{r},\ssm_r) = \frac{1}{9}\braket{m_{A,r},m_{B,r}|\big[V_{A,r} \otimes V_{B,r}\big]\rho_{\ssm_r}\big[V_{A,r}^{\dag} \otimes V_{B,r}^{\dag}\big]|m_{A,r},m_{B,r}},
\end{align}
where $\ket{m_{A,r},m_{B,r}}=\ket{m_{A,r}}\otimes \ket{m_{B,r}}$, and the prefactor $(\frac{1}{3})^2$ is the probability that we applied $V_{A,r}$ and $V_{B,r}$. Shadows have the defining property that \cite{huang2020predicting}
\begin{align}
    \sum_{V_{A,r} V_{B,r} m_{A,r} m_{B,r}} p(\rho^S_{r},\ssm_r) \rho^S_{r} = \rho_{m_r}.
\end{align}
This is to say that, if we could prepare $\rho_{\ssm}$ for a particular set of outcomes $\ssm$, and then extract many shadows of this particular density matrix, their average would be $\rho_{\ssm}$. This property of shadows allows for a straightforward construction of density matrices that can be prepared efficiently, but this is not the situation here. In the post-measurement setting, each state is $\rho_{\ssm}$ is typically prepared no more than once, so we cannot perform the average over shadows. From each repeat $r$ of the experiment we therefore extract $\ssm_r$ and a shadow $\rho^{\ssS}_r$ which can be expressed as
\begin{align}
    \rho^{\ssS}_{r} = \rho_{\ssm_r} + \eta_{r},
\end{align}
with $\sum_{V_{A,r} V_{B,r} m_{A,r} m_{B,r}}p(\rho^S_{r},\ssm)\eta_r=0$.

Consider now a weighted average of the observed shadows, with matrix-valued weights $w_{\ssm_r}$ depending on $\ssm_r$ only (and not $V_{A,r}$, $V_{B,r}$, $m_{A,r}$ or $m_{B,r}$),
\begin{align}
    \frac{1}{R}\sum_r w_{\ssm_r}\rho^{\ssS}_r = \frac{1}{R}\sum_{r} w_{\ssm_r}\rho_{m_r} +  \frac{1}{R}\sum_{r} w_{\ssm_r}\eta_r \xrightarrow[R \to \infty]{\phantom{a}} \sum_{\ssm} p_{\ssm} w_{\ssm} \rho_{\ssm}.\label{eq:Wav}
\end{align}
Because $w_{\ssm_r}$ depends only on $\ssm_r$, the average of $w_{\ssm_r}\eta_r$ over random unitary operations ($V_{A,r}$ and $V_{B,r}$) and probe outcomes ($m_{A,r}$ and $m_{B,r}$) is zero. This is simply the statement that $\sum_{V_{A,r} V_{B,r} m_{A,r} m_{B,r}}p(\rho^S_{r},\ssm_r)\eta_r=0$. Weighted averages over shadows therefore allow us to study properties of the ensemble of post-measurement states $\rho_{\ssm}$. Note that if we set $w_{\ssm}=1$, the average is $\sum_{\ssm}p_{\ssm}\rho_{\ssm}$; this is the state generated if we measure and discard the outcomes, i.e. it is the state generated by dephasing. Because a measurement with an unknown outcome can only have a local effect on our description of the state, the nonlocal phenomena that we aim to probe are invisible in $\sum_{\ssm}p_{\ssm}\rho_{\ssm}$.

As demonstrated in Ref.~\cite{garratt2024probing,mcginley2024postselection}, it is possible to construct weighted averages that provide rigorous bound on probes of post-measurement entanglement. The quantitites that we are interest in are the post-measurement von Neumann entropy $S_{\ssm}$ of $A$ and $B$, the coherent information $I_{\ssm}$ between $A$ and $B$, and negativity between $A$ and $B$. These quantities are defined by
\begin{align}
    S_{\ssm} = -\text{Tr}[\rho_{\ssm}\log_2 \rho_{\ssm}], \quad I_{\ssm} = S_{\ssm,A} - S_{\ssm}, \quad N_{\ssm} = -\text{Tr}[ \Pi(\rho^{\textsf{T}_A}_{\ssm}) \rho^{\textsf{T}_A}_{\ssm}].
\end{align}
Here $S_{\ssm,A}$ is the von Neumann entropy of the reduced density matrix $\rho_{\ssm,A}=\text{Tr}_B\rho_{\ssm}=\text{Tr}_B\rho_{\ssm,AB}$, $\rho^{\textsf{T}_A}_{\ssm}$ is the partial transpose of $\rho_{\ssm}$ with respect to degrees of freedom in $A$, and $\Pi(X)$ is the projector onto the span of the eigenstates of the matrix $X$ having negative eigenvalues. If $I_{\ssm}>0$ or $N_{\ssm}>0$, the state $\rho_{AB,\ssm}$ is non-separable (i.e. it is entangled).

Cross-correlations can be constructed to bound the measurement-averaged quantities 
$\sum_{\ssm}p_{\ssm}S_{\ssm}$, $\sum_{\ssm} p_{\ssm}I_{\ssm}$, $\sum_{\ssm} p_{\ssm} N_{\ssm}$. The basic idea is to choose weights $W_{\ssm}$ in Eq.~\eqref{eq:Wav} that are based on approximate computational models $\rho^{\ssC}_{\ssm}$ for the true (and inaccessible) post-measurement states $\rho_{\ssm}$. The models $\rho^{\ssC}_{\ssm}$ are valid density matrices. Choosing $W_{\ssm}=-\log \rho^{\ssC}_{\ssm}$ in Eq.~\eqref{eq:Wav} and taking a trace, we can determine the averages of
\begin{align}
    S^{\ssS \ssC}_r = -\text{Tr}\big[\rho^{\ssS}_{\ssm_r}\log_2 \rho^{\ssC}_{\ssm_r}], \quad S^{\ssQ\ssC}_{\ssm} = -\text{Tr}\big[\rho_{\ssm}\log_2 \rho^{\ssC}_{\ssm}],
\end{align}
it can be seen that $\frac{1}{R}\sum_r S^{\ssS\ssC}_r$ converges to $\sum_{\ssm}p_{\ssm}S^{\ssQ\ssC}_{\ssm}$ in the limit of large $R$. The quantity $S^{\ssS\ssC}_r$ can be determined for each $r$ from the observed shadow $\rho^{\ssS}_{r}$ and a computational model $\rho^{\ssC}_{\ssm_r}$. Then, observing that the quantum Kullback-Leibler divergence $D_{\ssm}$ between $\rho_{\ssm}$ and $\rho_{\ssm}^{\ssC}$ can be expressed as $D_{\ssm} = S^{\ssQ\ssC}_{\ssm}-S_{\ssm}$, we immediately have $S^{\ssQ\ssC}_{\ssm} \geq S_{\ssm}$, simply because the quantum Kullback-Leibler divergence is non-negative \cite{nielsen2010quantum}. Therefore, at large $R$,
\begin{align}
   \frac{1}{R} \sum_r S^{\ssS \ssC}_r + \epsilon_S \geq \sum_{\ssm} p_{\ssm} S_{\ssm},
\end{align}
which defines $\epsilon_S$, the difference between the true average and the average obtained from $R$ repeats. 
At finite $R$ we have $\epsilon_S \sim \pm R^{-1/2}$, with a prefactor that is of order unity when the number of probe qubits is of order unity. Note that the above inequality is satisifed for any density matrix $\rho^{\ssC}_{\ssm}$, so it provides an objective characterization of the ensemble of post-measurement states. The inequality is saturated when $\rho_{\ssm}=\rho_{\ssm}^{\ssC}$.

To arrive at the lower bound on measurement-averaged coherent information, we define 
\begin{align}
    I^{\ssS\ssC}_{r} = S^{\ssS \ssC}_{r,A} - S^{\ssS \ssC}_r, \quad I^{\ssQ \ssC}_{\ssm} = S^{\ssQ\ssC}_{\ssm,A} - S^{\ssQ\ssC}_{\ssm},
\end{align}
where
$S^{\ssS \ssC}_{r,A} = -\text{Tr}\big[\rho^{\ssS}_{r,A}\log \rho^{\ssC}_{\ssm_r,A}]$ and $S^{\ssQ \ssC}_{\ssm,A} = -\text{Tr}\big[\rho_{\ssm,A}\log \rho^{\ssC}_{\ssm,A}]$. The average $\frac{1}{R}\sum_r I^{\ssS\ssC}_r$ converges to $\sum_{\ssm}p_{\ssm}I^{\ssQ\ssC}_{\ssm}$ at large $R$. The lower bound follows from the fact that the quantum Kullback-Leibler divergence is non-increasing under quantum channels. Since tracing out a subsystem can be expressed as a quantum channel, we have $D_{\ssm} \geq D_{\ssm,A}$, i.e.
\begin{align}
    S^{\ssQ\ssC}_{\ssm}-S_{\ssm} \geq S^{\ssQ\ssC}_{\ssm,A}-S_{\ssm,A},
\end{align}
where the right-hand side of this inequality is $D_{\ssm,A}$. Rearranging this inequality, we arrive at $I^{\ssQ\ssC}_{\ssm} \leq I_{\ssm}$. Again, this inequality is saturated for a perfect model. We then have
\begin{align}
   \frac{1}{R} \sum_r I^{\ssS \ssC}_r + \epsilon_I \leq \sum_{\ssm} p_{\ssm} I_{\ssm},
\end{align}
where again the error $\epsilon_I$ in our estimate for the mean decays as $\epsilon_I \sim \pm R^{-1/2}$ with a prefactor of order unity (in the case where the number of probe qubits is of order unity). Therefore, if the mean $I^{\ssS \ssC}_r > 0$, and the mean is much larger than the standard error in the mean, our cross-correlations can show that $\sum_{\ssm} p_{\ssm} I_{\ssm}>0$ with high probability, and therefore that the post-measurement states are entangled.

We can make similar statements based on the measurement-averaged negativity. In this case we define
\begin{align}
    N^{\ssS\ssC}_{r} = -\text{Tr}\big[ (\rho^S_r)^{\textsf{T}_A} \Pi\big( (\rho^{\ssC}_{\ssm_r})^{\textsf{T}_A}\big)\big], \quad N^{\ssQ\ssC}_{\ssm} = -\text{Tr}\big[ (\rho_{\ssm})^{\textsf{T}_A} \Pi\big( (\rho^{\ssC}_{\ssm})^{\textsf{T}_A}\big)\big].
\end{align}
At large $R$, the average $\frac{1}{R}\sum_r N^{\ssS\ssC}_{r}$ converges to $\sum_{\ssm}p_{\ssm} N^{\ssQ \ssC}_{\ssm}$, and we denote the fluctuations in our average around the true mean at finite $R$ by $\epsilon_N \sim \pm R^{-1/2}$. To arrive at the bound, focusing for now on a specific $\ssm$, it is convenient to write the spectral decompositions of the partial transposed density matrix as $\rho^{\textsf{T}_A}_{\ssm}=\sum_{\nu}\lambda_{\nu}\ket{\nu}\bra{\nu}$, and similar for $(\rho^{\ssC}_{\ssm})^{\textsf{T}_A}=\sum_{\nu}\lambda^{\ssC}_{\nu}\ket{\nu^{\ssC}}\bra{\nu^{\ssC}}$. These spectral decompositions depend on $\ssm$, but we suppress their $\ssm$ dependence for brevity. These matrices are Hermitian, so all eigenvalues are real and the eigenstates are orthonormal, but they are not necessarily valid density matrices, so the eigenvalues can be negative. In terms of the spectral decomposition,
\begin{align}
    N^{\ssQ\ssC}_{\ssm} = \sum_{\mu| \lambda^{\ssC}_{\mu}<0}\sum_{\nu|\lambda_{\nu} < 0}|\lambda_{\nu}| |\braket{\mu^{\ssC}|\nu}|^2 - \sum_{\mu| \lambda^{\ssC}_{\mu}<0}\sum_{\nu|\lambda_{\nu} > 0}\lambda_{\nu} |\braket{\mu^{\ssC}|\nu}|^2 \leq \sum_{\mu}\sum_{\nu|\lambda_{\nu} < 0}|\lambda_{\nu}||\braket{\mu^{\ssC}|\nu}|^2 = N_{\ssm}.
\end{align}
The first equality follows from dividing the sum over eigenstates $\ket{\nu}$ of $\rho_{\ssm}$ into contributions with $\lambda_{\nu}<0$ and $\lambda_{\nu}>0$. The second term in the resulting expression is positive, while the first is upper bounded by a sum involving all $\mu$ (rather than just those with $\lambda_{\ssm}^{\ssC}<0$). The final equality follows from the sum over $\mu$. Again, the resulting inequality $N^{\ssQ\ssC}_{\ssm} \leq N_{\ssm}$ is saturated when $\rho^{\ssC}_{\ssm}=\rho_{\ssm}$. Therefore
\begin{align}
   \frac{1}{R} \sum_r N^{\ssS \ssC}_r + \epsilon_N \leq \sum_{\ssm} p_{\ssm} N_{\ssm},
\end{align}
so at large $R$ where the flucutations $\epsilon_N$ are small we arrive at a lower bound on the measurement-averaged negativity in terms of cross-correlations. If we find that the $\frac{1}{R} \sum_r N^{\ssS \ssC}_r$ is greater than zero and much larger than the fluctuations, we can show that $\sum_{\ssm} p_{\ssm} N_{\ssm}>0$ with high probability.

\section{Additional experimental results on one-dimensional cluster states}\label{sec:1dextra}

In this section we first experimentally probe the coherent information in one-dimensional cluster states, using the same computational models as in the main text. Following this we consider an alternative way to detect measurement-induced phenomena in experiments; this involves classifying observations using computational models, rather than cross-correlating our observations with such models. Recall that, in our experiments on one-dimensional cluster states, the probe qubits sit at the two ends of a chain with open boundary conditions, and we aim to probe entanglement between these qubits that is induced by measurements of all others.

\subsection{Coherent information}
In the main text we detected measurement-induced entanglement using a lower bound $\overline{N^{\ssQ\ssC}_{\ssm}}$ on the average entanglement negativity $\overline{N}_{\ssm}$. Another probe of mixed-state entanglement is the coherent information. The post-measurement coherent information is of significant interest in the context of quantum error correction, where it quantities the amount of quantum information that is recoverable following an error channel and the measurement of a large number of syndromes \cite{schumacher1996quantum}. 

\begin{figure}
\includegraphics[width=\textwidth]{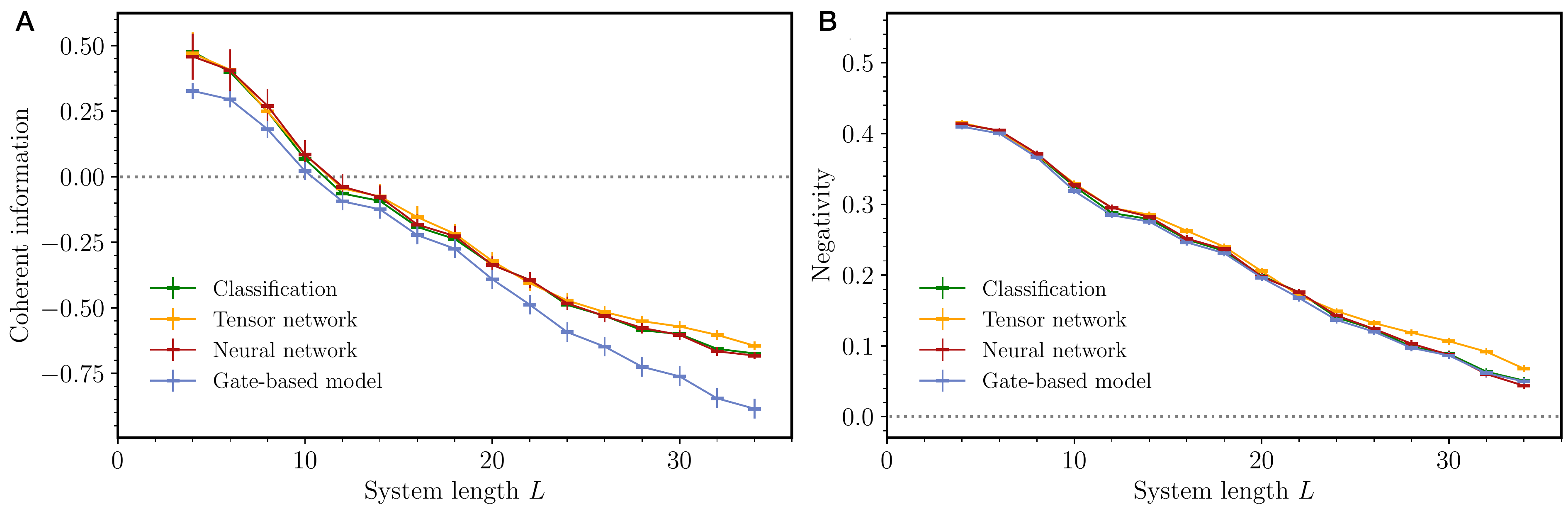}
\caption{Measurement-induced mixed-state entanglement in one-dimensional cluster states. As in the main text, the probe qubits are at the two ends of a chain with open boundary conditions. Estimates for  the average (\textsf{A}) coherent information $\overline{I_{\ssm}}$ and (\textsf{B})  negativity $\overline{N_{\ssm}}$ are measured using the classification method described in Sec.~\ref{sec:classification} (green), and lower bounds for these quantities are constructed using cross-correlations as described in the main text and in Sec.~\ref{sec:QC}. These cross-correlations involve estimates $\rho^{\ssC}_{\ssm}$ for post-measurement density matrices that are generated using optimized matrix product states (orange, see Sec.~\ref{subsec:mpo}), attention-based neural networks (red, see Sec.~\ref{subsec:TranNN}) and explicit calculations based on knowledge of gates (blue). As in the main text, in the gate-based model we initially do not include noise, and so generate a pure estimate $\hat \rho^{\ssC}_{\ssm}$ for the post-measurement density matrix. From $\hat \rho^{\ssC}_{\ssm}$ we construct the mixed state $\rho^{\ssC}_{\ssm}=\epsilon \hat \rho^{\ssC}_{\ssm} + (\epsilon/4)\mathds{1}$ ($\epsilon=0.3$) that is then used in cross-correlations. Note that, aside from the results using the classification method, the data in panel \textsf{B} appears in Fig.~2 of the main text.}
\label{fig:1dextra}
\end{figure}

With this application in mind, it is valuable to have a general scheme to lower bound the measurement-averaged coherent information $\overline{I_{\ssm}}$. As a proof-of-principle, here we measure various lower bounds $\overline{I^{\ssQ\ssC}_{\ssm}}$ on the coherent information between probe qubits $A$ and $B$ in the one-dimensional cluster state. The results are shown in Fig.~\ref{fig:1dextra}\textsf{A}: the red and orange data is obtained using the same unsupervised machine learning models for $\rho^{\ssC}_{\ssm}$ as in the main text, while the blue data is obtained using the gate-based model. It is clear that the unsupervised models can detect post-measurement coherent information up to $L = 10$, while the lower bound drops below zero for $L \geq 11$. 

It is useful to compare the lower bounds on the coherent information with the lower bounds on negativity. In both cases, a positive lower bound implies that the post-measurement states are entangled. The fact that $\overline{I^{\ssQ\ssC}_{\ssm}}$drops below zero for $L \geq 11$ simply means that this quantity does not detect the entanglement that we know is present from our measurements of $\overline{N^{\ssQ\ssC}_{\ssm}}$ in Fig.~2 of the main text (these are reproduced in Fig.~\ref{fig:1dextra}\textsf{B}).

\subsection{Classification by simulation}\label{sec:classification}

We have shown that cross-correlations between computational models $\ssm \mapsto \rho^{\ssC}_{\ssm}$ and observed shadows $\rho^{\ssS}_{\ssm}$ provide a way to detect measurement-induced entanglement. A related alternative, which we explore here, is to `bin', or classify, observed shadows according to a computational model for the system.  

The general idea is as follows. For post-measurement states of $K$ qubits (here $K=2$), we can partition the space of valid $K$-qubit density matrices into disjoint classes $\textsf{M}_i$. Here the index $i$ labels these classes. Given a model which maps sets of measurement outcomes $\ssm$ to estimates for post-measurement states $\rho^{\ssC}_{\ssm}$, if the estimate $\rho^{\ssC}_{\ssm} \in \textsf{M}_i$ we say that $\ssm \in \textsf{M}_i$. After $R$ repeats of the experiment ($r=1,\ldots,R$) we have $R$ sets of outcomes $\ssm_r$ and $R$ shadows $\rho^{\ssS}_{r}$ of the post-measurement state. For each $i$ we then divide the class of outcomes $\ssm \in \textsf{M}_i$ into two disjoint sets, of size $R_1$ and $R_2$ with $R_1+R_2=R$ (these numbers are $i$-dependent, but we suppress this dependence for brevity), and construct
\begin{align}
    \rho_{i1} = \mathds{E}_{r \in R_1, \,\ssm_r \in \textsf{M}_i} \rho^{\ssS}_r, \quad  \rho_{i2} = \mathds{E}_{r \in R_1, \, \ssm_r \in \textsf{M}_i} \rho^{\ssS}_r,
\end{align} 
where e.g. $\mathds{E}_{r \in R_1, \,\ssm_r \in \textsf{M}_i}$ denotes an average over repeats of the experiment with outcome $\ssm_r$ in class $\textsf{M}_i$ and set $R_1$. The matrices $\rho_{i1}$ and $\rho_{i2}$ are averages over shadows observed in runs with $\ssm \in \textsf{M}_i$. Note after a finite number of repeats these matrices are not necessarily positive semidefinite; to resolve this, in classical post-processing one can `depolarize' $\rho_{i2}$, i.e. we replace $\rho_{i2} \to (1-\epsilon)\rho_{i2}+(\epsilon/2^K)\mathds{1}$, with $\epsilon$ chosen so that $\rho_{i2}$ is positive semidefinite. In our implementation below, however, we find that this is not necessary. In analogy with quantum-classical cross-correlations, we can then construct the quantities 
\begin{align}
    S_i = -\text{Tr}[\rho_{i1}\log_2 \rho_{i2}], \quad I_i = S_{i,A}-S_i, \quad N_i = -\text{Tr}\big[ \rho_i^{\mathsf{T}_A}\Pi\big(\rho_i^{\mathsf{T}_A}\big)].
\end{align}
In the limit of a large number of repeats of the experiment, where the parameter $\epsilon$ necessary to make $\rho_{i2}$ positive semidefinite approaches zero, the above quantities converge to properties of the density matrix $\sum_{\ssm \in \textsf{M}_i}p_{\ssm}\rho_{\ssm}$, where $p_{\ssm}$ is the Born probability associated with the outcomes $\ssm$.

For the one-dimensional array, in the absence of noise, given measurement outcomes $\ssm$ on the central qubits (i.e. those not at the ends of the chain), there are only four possible post-measurement states $\rho_{\ssm}$ of the probes. These are the four standard maximally entangled two-qubit states (discussed in Methods). By classifying post-measurement states in this way, we measure $I_i$ and $N_i$ using the method described above. Our results are shown in Fig.~\ref{fig:1dextra}, and we compare the results with cross-correlations constructed from various models $\ssm \mapsto \rho^{\ssC}_{\ssm}$.

In Figs.~\ref{fig:1dextra}\textsf{A} and \textsf{B} we show estimates for the average coherent information and negativity, respectively, obtained using the classification approach described above. There we find comparable results to our lower bounds based on cross-correlations between computational models and experimental data.  

\section{Dependence of learning dynamics on sample number and training time}

Here we study the impact of increasing the number of samples $n$ and epochs $t$ used in training in our experiments on two-dimensional arrays. Recall that in the main text we showed that, on increasing $t$ for a fixed $n=n_{\rm{max}}$, the form of the decrease of $\overline{S^{\ssQ \ssC}_{\ssm}}$ is rather complicated. On the other hand, for fixed $t$, we found a clear logarithmic decay on increasing $n$. 

\begin{figure}
\includegraphics[width=\textwidth]{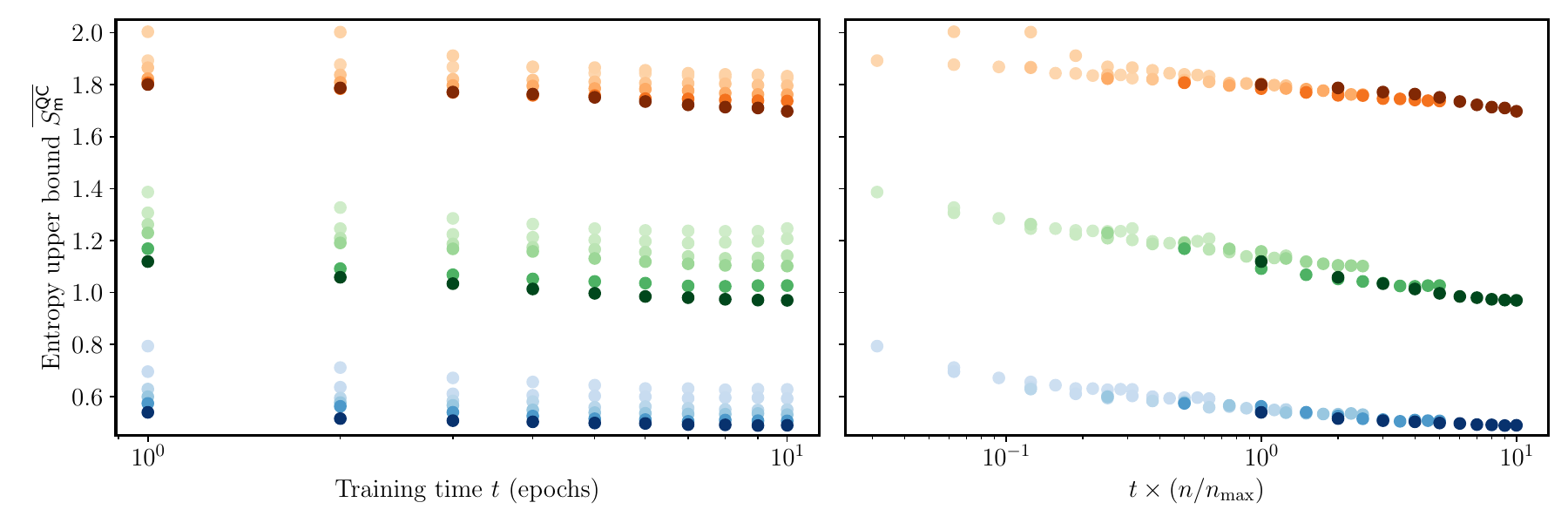}
\caption{Dependence of the upper bound on the average post-measurement von Neumann entropy $\overline{S^{\ssQ \ssC}_{\ssm}}$ (for the two probe qubits in the two-dimensional array) on the number of samples $n$ and epochs $t$ used in training. Here we show $\theta =0.15$ (blues), $\theta=0.25$ (greens) and $\theta=0.35$ (oranges) for $n/n_{\rm{max}} = 1/32, 1/64, 1/16, 1/8, 1/4, 1/2, 1$ (light to dark), where $n_{\rm{max}}=7.8\times 10^7$. The upper bound $\overline{S^{\ssQ \ssC}_{\ssm}}$ is computed from cross-correlations between NN predictions and shadows obtained from a separate $10^6$ repeats of the experiment, as in the main text. \textsf{A}: Increasing $t$ initially leads to a decrease in the upper bound, but overfitting becomes apparent beyond a $t$ that is small for small data fractions $n/n_{\rm{max}}$. \textsf{B}: Collapse of $\overline{S^{\ssQ \ssC}_{\ssm}}$ versus $t \times n/n_{\rm{max}}$, for various $n/n_{\rm{max}}$ displayed also in \textsf{A}. The quantity $t \times n/n_{\rm{max}}$ is proportional to the number of updates of the NN parameters in training.}
\label{fig:collapse}
\end{figure}

In Fig.~\ref{fig:collapse}\textsf{A} we show $\overline{S^{\ssQ \ssC}_{\ssm}}$ versus $t$ for different numbers $n$ of samples, and for $\theta=0.15$, $0.25$, and $0.35$. When we use a small fraction $n/n_{\rm{max}}$ of the full dataset our results indicate overfitting: we eventually find \emph{increases} in $\overline{S^{\ssQ \ssC}_{\ssm}}$ at large $t$. This is part of the reason why $\overline{S^{\ssQ \ssC}_{\ssm}}$ has a complicated dependence on $n$ at fixed $t$. 

During each epoch the NN accesses each of the $n \leq n_{\rm{max}} = 7.8 \times 10^7$ samples, so the total computational time for training is proportional to $nt$ rather than just $t$. The product $nt$ is therefore also proportional to the number of updates of NN parameters throughout training. It is therefore natural to ask how $\overline{S^{\ssQ \ssC}_{\ssm}}$ depends on $nt$, and in Fig.~\ref{fig:collapse} we find a striking collapse of the data for various $n/n_{\rm{max}}$. For each value of $\theta$, and crucuially for $t$ well below the point where we find evidence of overfitting, our data is well-described by
\begin{align}
    \overline{S^{\ssQ \ssC}_{\ssm}} = a - b \log_2 (nt), \label{eq:learningrate}
\end{align}
where each of $a$ and $b$ depend on $\theta$ as well as the system size. 

In the limit where the NN is large (i.e. where one is not limited in classical memory), we expect that for $n$ and $t$ that are no larger than exponential in the number $L^2$ of qubits, $\overline{S^{\ssQ \ssC}_{\ssm}}$ should convert to $\overline{S_{\ssm}}$. If this is the case, then the logarithmic rate $b$ cannot decay with $L$ faster than $L^{-2}$. Because of this, it is possible to determine $b$ with high accuracy using a number of experimental repeats (and computational time) that is no larger than polynomial in $L$. A learnability transition can therefore be observed from the dependence of $b$ on $\theta$. While in the present work we have accessed a limited range of system sizes, testing Eq.~\eqref{eq:learningrate} is an interesting problem for the future.

It is important to note that extracting this probe of the learnability transition does not imply the ability to detect measurement-induced entanglement. For example, in the highly entangled phase we have seen that $a \approx 2$, which is the value for maximally mixed two-qubit $\rho^{\ssC}_{\ssm}$. Constructing a nontrivial lower bound on negativity requires a model $\rho^{\ssC}_{\ssm}$ that does not have a positive partial transpose, but such states are a finite distance from the maximally mixed state. We therefore expect to require $\overline{S^{\ssQ \ssC}_{\ssm}} < 2 - \delta$, for $\delta$ of order unity, before we can detect negativity. If $b$ is inversely proportional to the number of qubits, then although we can efficiently determine its value, achieving $\overline{S^{\ssQ \ssC}_{\ssm}} < 2 - \delta$ would require $nt$ that is exponential in system size. 

\section{Nonlocal effects of measurements in two-dimensional arrays}\label{sec:2dextra}

Post-measurement states of the probe qubits can depend sensitively on measurement outcomes observed on the non-probe qubits. In one-dimensional cluster states, with probe qubits at the two ends of the chain, in the absence of noise we know that flipping a single measurement outcome on a non-probe qubit $k$, i.e. $\ssm \to \tilde\ssm$ with $m_j = \tilde m_j$ for all $j \neq k$ and $m_k \neq \tilde m_k$, causes a drastic change in the post-measurement state, with $\rho_{\ssm}$ and $\rho_{\tilde \ssm}$ orthogonal. In two-dimensional cluster states the dependence of the post-measurement states on measurement basis and outcomes is less obvious. Here we show how to study this dependence using our trained neural network $\ssm \mapsto \rho^{\ssC}_{\ssm}$ and the experimental data. We also investigate the distribution of observed measurement outcomes $\ssm$.

\subsection{Sensitivity of post-measurement states to distant outcomes}

\begin{figure}
\includegraphics[width=\textwidth]{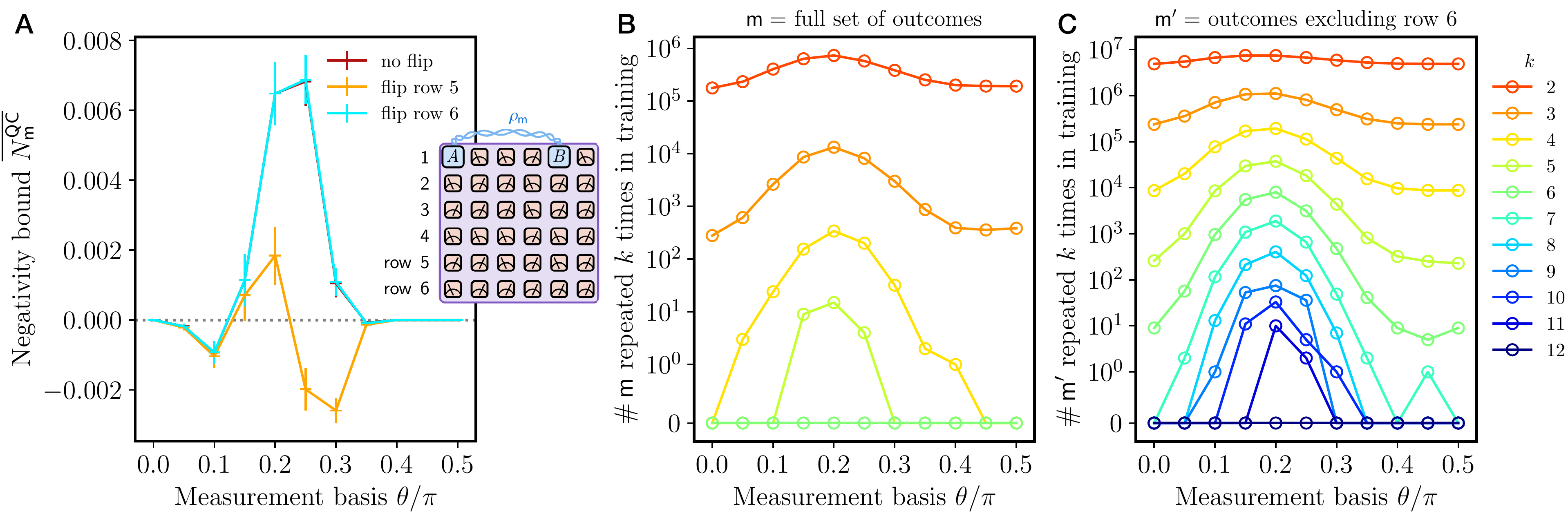}
\caption{Complexity of learning post-measurement entanglement in cluster states of $6 \times 6$ arrays. Here probe qubits are separated by distance $d=4$ in row $1$ (see diagram), as in Fig.~4 of the main text. \textsf{A}: Significance of distant measurement outcomes in detection of post-measurement entanglement. Here we show the lower bound $\overline{N^{\ssQ\ssC}_{\ssm}}$ on average negativity measured using the attention-based neural network $\ssm \mapsto \rho^{\ssC}_{\ssm}$ (red, also shown in Fig.~4), and study the effect of modifying the input to the neural network. We show lower bounds obtained from cross-correlations between shadows $\rho^{\ssS}_{\ssm}$ and $\rho^{\ssC}_{\tilde\ssm}$, with $\tilde\ssm$ obtained by flipping all outcomes $m$ in row $5$ (orange) and row $6$ (cyan). The negativity is highly sensitive to outcomes in row $5$ but not to those in row $6$. \textsf{B}: Number of repeats of the different possibility outcomes $\ssm$ observed on the $34$ non-probe qubits in training, corresponding to $7.8\times 10^7$ repeats of the experiment. No set of outcomes $\ssm$ is observed more than five times. At intermediate $\theta$ the distribution of Born probabilities $p_{\ssm}$ broadens, making some outcomes more likely. \textsf{C}: Number of repeats of different possible outcomes $\ssm'$ observed on the $28$ non-probe qubits that are \textit{not} in row $6$. Even excluding row $6$, no set of outcomes $\ssm'$ is observed more than $11$ times.}
\label{fig:negativitysensitivity}
\end{figure}

Let us first establish our notation. For two-dimensional cluster states of $6 \times 6$ qubit arrays we label the rows $1,2,\ldots,6$, with the two probe qubits $A$ and $B$ separated by a distance $d=4$ in row $1$. This notation is indicated in the diagram in Fig.~\ref{fig:negativitysensitivity}\textsf{A}. After training the neural network $\ssm \mapsto \rho^{\ssC}_{\ssm}$ on $7.8 \times 10^7$ repeats of the experiment, we have shown in the main text that cross-correlations between this model and the observed shadows $\rho^{\ssS}_{\ssm}$ can detect entanglement between $A$ and $B$. The successful detection of entanglement between $A$ and $B$ suggests that $\rho^{\ssC}_{\ssm}$ is a reasonable approximation to the true post-measurement state $\rho_{\ssm}$, or at least that the projectors $\Pi([\rho^{\ssC}_{\ssm}]^{\mathsf{T}_A})$ and $\Pi([\rho_{\ssm}]^{\mathsf{T}_A})$ onto the negative eigenspaces of the partial transposes of these density matrices are in good agreement.

We can ask how the post-measurement state depends on distant outcomes by studying the effect of flipping distant measurement outcomes; here we flip all outcomes $m_j$ in row $5$, or all outcomes in row $6$. We denote by $\tilde \ssm$ the set of outcomes obtained from $\ssm$ through such a flip. If the true post-measurement states depend on the outcomes that we flip, we will find a significant difference between our lower bound $\overline{N^{\ssQ\ssC}_{\ssm}} = -\overline{\text{Tr}[\rho^{\ssS}_{\ssm}\Pi([\rho^{\ssC}_{\ssm}]^{\mathsf{T}_A})]}$ and $-\overline{\text{Tr}[\rho^{\ssS}_{\ssm}\Pi([\rho^{\ssC}_{\tilde\ssm}]^{\mathsf{T}_A})]}$. On the other hand, if these quantities are close to another, this implies that the post-measurement state is not sensitive to the flip $\ssm \to \tilde \ssm$. 

The results of these experiments are shown in Fig.~\ref{fig:negativitysensitivity}\textsf{A}. There we show that when $\tilde \ssm$ is obtained by flipping the outcomes in $\ssm$ on row $6$, the lower bound on the negativity is barely changed. Therefore, at intermediate $\theta$ where we observe the peak in post-measurement negativity, $\rho^{\ssC}_{\ssm}$ has almost no dependence on outcomes $m_j$ in row $6$. By contrast, flipping the outcomes in row $5$ has a dramatic effect on the detected negativity. For example, for $\theta/\pi=0.25$ and $0.3$ the model prompted with the correct outcomes $\ssm$ detects entanglement, while the model prompted with the incorrect outcomes $\tilde \ssm$ does not. These results show clearly that that the structure of the post-measurement state has a highly nonlocal dependence on outcomes across the array.

\subsection{Statistics of measurement outcomes}
Since at intermediate $\theta$ we can detect entanglement with a model $\rho^{\ssC}_{\ssm}$ that is insensitive to the outcomes in row $6$, it is natural to ask whether an alternative approach based on postselection is feasible. Let us denote by $\ssm'$ the set of outcomes $m_j$ excluding row $6$, i.e. $\ssm$ is a string of $34$ outcomes while $\ssm'$ is a string of $28$ outcomes. By ignoring row $6$, is it possible that the neural network simply sees the same $\ssm'$ so many times that it can `learn' to reconstruct a good estimate for the true $\rho_{\ssm}$? In other words, has the experiment been repeated so many times that we could have just ignored row $6$, and perfomed quantum state tomography on $\rho_{\ssm}$? Since $\rho_{\ssm}$ is a state of two qubits, it has $15$ real parameters. Tomography of a particular $\rho_{\ssm}$ therefore requires that it is created far more than $15$ times.

In Fig.~\ref{fig:negativitysensitivity}\textsf{B} we first show the number of times different sets of outcomes $\ssm$ are observed in the $7.8 \times 10^7$ runs of the experiment used for training the neural network. If all outcomes were equally likely, occurring with probability $2^{-34}$ each, the expected number of outcomes $\ssm$ observed exactly twice would be $\approx 1.7 \times 10^5$. Although this is the same order of magnitude as the $k=2$ results in Fig.~\ref{fig:negativitysensitivity}\textsf{B}, the difference simply reflects that fact that the true distribution of Born probabilities $p_{\ssm}$ is broad. Interestingly, the peak observed in the number of repetitions at intermediate $\theta$ indicates a broadening of the distribution for the same measurement bases where we are able to detect negativity. It is nevertheless the case that no set of outcomes $\ssm$ is observed more than five times.

Next, in Fig.~\ref{fig:negativitysensitivity}\textsf{C}, we restrict our attention to rows $1$ through $5$, neglecting row $6$. Our results show that no set of outcomes $\ssm'$ is observed more than $12$ times, and we do not expect that accurate tomography of any one of the post-measurement states would be possible with so few observations. Moreover, the vast majority of the $\ssm'$ observed during training are observed only once (i.e. for $k=1$ we have at least $6.9 \times 10^7$ non-repeating outcomes for each $\theta$).

Although our discussion above has focused on the actual number of repeats of different sets of measurement outcomes in a $L \times L$ array with $L=6$, it is important to recall that schemes based on postselection are not scalable to large systems. The results in Fig.~\ref{fig:negativitysensitivity}\textsf{A} indicate that the post-measurement states are sensitive to distant outcomes, so the number of repeats necessary to generate a particular $\rho_{\ssm}$ is exponentially small in $L^2$. By contrast, at intermediate $\theta$ our results suggest that machine learning provides a scalable way to detect measurement-induced entanglement.

\section{Numerical characterization of gate-based models}\label{sec:noiseless2}

In the main text we have used gate-based models to construct lower bounds on measurement-induced entanglement. Here we numerically investigate the pure post-measurement states $\hat\rho^{\ssC}_{\ssm}$ generated from two-dimensional cluster cluster states. We numerically sample measurement outcomes $\ssm$, and compute averages of the post-measurement negativity over the sampled $\ssm$. 

\begin{figure}[htbp]
\begin{center}
\includegraphics[width=\textwidth]{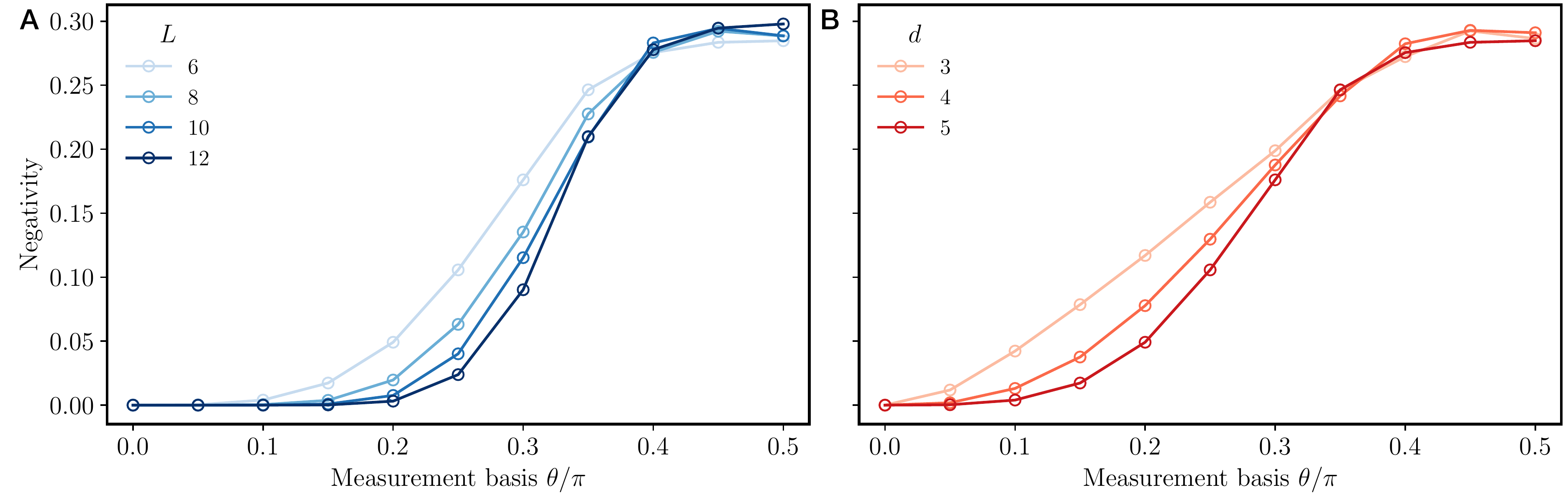}
\caption{Numerical calculations of average entanglement negativity between probe qubits $A$ and $B$ in pure post-measurement states $\hat \rho_{\ssm}$. The states $\hat \rho_{\ssm}$ are determined from error-free models of the gates used to prepare two-dimensional cluster states. \textsf{A}: Probe qubits at edge-sharing corners of $L\times L$ cluster states, for various $L$ (legend). \textsf{B}: Probe qubits at various separations $d$ (legend) along an edge of a $6 \times 6$ array. The specific locations of probe qubits are indicated in the diagram in Fig. 4 of the main text. Here we average over $10^6$ samples of the outcomes $\ssm$, so the standard error in the displayed data is of order $10^{-3}$ (error bars are not shown).}
\label{fig:simulation}
\end{center}
\end{figure}

In Fig.~\ref{fig:simulation}\textsf{A} we calculate the measurement-averaged negativity of the density matrices $\rho^{\ssC}_{\ssm}$ in the case where the probe qubits are at two edge-sharing corners of a square array of $L \times L$ qubits, for various $L$. On increasing $\theta$ we see that the negativity increases, and on increasing $L$ the onset of entanglement between probe qubits becomes sharper. This behavior, in particular the crossing around $\theta/\pi=0.4$, is indicative of an entanglement transition. In Fig.~\ref{fig:simulation}\textsf{B} we simulate the $6 \times 6$ array studied in the main text with probe qubits at various separations $d$ (in the locations illustrated in Fig. 4 of the main text). For large $\theta$ the negativity between the probes becomes approximately independent of $d$.

\begin{thebibliography}{56}%
\makeatletter
\providecommand \@ifxundefined [1]{%
 \@ifx{#1\undefined}
}%
\providecommand \@ifnum [1]{%
 \ifnum #1\expandafter \@firstoftwo
 \else \expandafter \@secondoftwo
 \fi
}%
\providecommand \@ifx [1]{%
 \ifx #1\expandafter \@firstoftwo
 \else \expandafter \@secondoftwo
 \fi
}%
\providecommand \natexlab [1]{#1}%
\providecommand \enquote  [1]{``#1''}%
\providecommand \bibnamefont  [1]{#1}%
\providecommand \bibfnamefont [1]{#1}%
\providecommand \citenamefont [1]{#1}%
\providecommand \href@noop [0]{\@secondoftwo}%
\providecommand \href [0]{\begingroup \@sanitize@url \@href}%
\providecommand \@href[1]{\@@startlink{#1}\@@href}%
\providecommand \@@href[1]{\endgroup#1\@@endlink}%
\providecommand \@sanitize@url [0]{\catcode `\\12\catcode `\$12\catcode
  `\&12\catcode `\#12\catcode `\^12\catcode `\_12\catcode `\%12\relax}%
\providecommand \@@startlink[1]{}%
\providecommand \@@endlink[0]{}%
\providecommand \url  [0]{\begingroup\@sanitize@url \@url }%
\providecommand \@url [1]{\endgroup\@href {#1}{\urlprefix }}%
\providecommand \urlprefix  [0]{URL }%
\providecommand \Eprint [0]{\href }%
\providecommand \doibase [0]{https://doi.org/}%
\providecommand \selectlanguage [0]{\@gobble}%
\providecommand \bibinfo  [0]{\@secondoftwo}%
\providecommand \bibfield  [0]{\@secondoftwo}%
\providecommand \translation [1]{[#1]}%
\providecommand \BibitemOpen [0]{}%
\providecommand \bibitemStop [0]{}%
\providecommand \bibitemNoStop [0]{.\EOS\space}%
\providecommand \EOS [0]{\spacefactor3000\relax}%
\providecommand \BibitemShut  [1]{\csname bibitem#1\endcsname}%
\let\auto@bib@innerbib\@empty
\bibitem [{\citenamefont {Gross}\ \emph {et~al.}(2009)\citenamefont {Gross},
  \citenamefont {Flammia},\ and\ \citenamefont {Eisert}}]{gross2009most}%
  \BibitemOpen
  \bibfield  {author} {\bibinfo {author} {\bibfnamefont {D.}~\bibnamefont
  {Gross}}, \bibinfo {author} {\bibfnamefont {S.~T.}\ \bibnamefont {Flammia}},\
  and\ \bibinfo {author} {\bibfnamefont {J.}~\bibnamefont {Eisert}},\
  }\bibfield  {title} {\bibinfo {title} {Most quantum states are too entangled
  to be useful as computational resources},\ }\href
  {https://doi.org/10.1103/PhysRevLett.102.190501} {\bibfield  {journal}
  {\bibinfo  {journal} {Phys. Rev. Lett.}\ }\textbf {\bibinfo {volume} {102}},\
  \bibinfo {pages} {190501} (\bibinfo {year} {2009})}\BibitemShut {NoStop}%
\bibitem [{\citenamefont {Abrams}\ and\ \citenamefont
  {Lloyd}(1998)}]{abrams1998nonlinear}%
  \BibitemOpen
  \bibfield  {author} {\bibinfo {author} {\bibfnamefont {D.~S.}\ \bibnamefont
  {Abrams}}\ and\ \bibinfo {author} {\bibfnamefont {S.}~\bibnamefont {Lloyd}},\
  }\bibfield  {title} {\bibinfo {title} {Nonlinear quantum mechanics implies
  polynomial-time solution for $\mathrm{NP}$-complete and $\#\mathrm{P}$
  problems},\ }\href {https://link.aps.org/doi/10.1103/PhysRevLett.81.3992}
  {\bibfield  {journal} {\bibinfo  {journal} {Phys. Rev. Lett.}\ }\textbf
  {\bibinfo {volume} {81}},\ \bibinfo {pages} {3992} (\bibinfo {year}
  {1998})}\BibitemShut {NoStop}%
\bibitem [{\citenamefont {Aaronson}(2005)}]{aaronson2005quantum}%
  \BibitemOpen
  \bibfield  {author} {\bibinfo {author} {\bibfnamefont {S.}~\bibnamefont
  {Aaronson}},\ }\bibfield  {title} {\bibinfo {title} {Quantum computing,
  postselection, and probabilistic polynomial-time},\ }\href
  {http://doi.org/10.1098/rspa.2005.1546} {\bibfield  {journal} {\bibinfo
  {journal} {Proc. R. Soc. A}\ }\textbf {\bibinfo {volume} {461}},\ \bibinfo
  {pages} {3473} (\bibinfo {year} {2005})}\BibitemShut {NoStop}%
\bibitem [{\citenamefont {Bremner}\ \emph {et~al.}(2011)\citenamefont
  {Bremner}, \citenamefont {Jozsa},\ and\ \citenamefont
  {Shepherd}}]{bremner2011classical}%
  \BibitemOpen
  \bibfield  {author} {\bibinfo {author} {\bibfnamefont {M.~J.}\ \bibnamefont
  {Bremner}}, \bibinfo {author} {\bibfnamefont {R.}~\bibnamefont {Jozsa}},\
  and\ \bibinfo {author} {\bibfnamefont {D.~J.}\ \bibnamefont {Shepherd}},\
  }\bibfield  {title} {\bibinfo {title} {Classical simulation of commuting
  quantum computations implies collapse of the polynomial hierarchy},\ }\href
  {https://doi.org/https://doi.org/10.1098/rspa.2010.0301} {\bibfield
  {journal} {\bibinfo  {journal} {Proc. R. Soc. A}\ }\textbf {\bibinfo {volume}
  {467}},\ \bibinfo {pages} {459} (\bibinfo {year} {2011})}\BibitemShut
  {NoStop}%
\bibitem [{\citenamefont {Bell}(1964)}]{bell1964einstein}%
  \BibitemOpen
  \bibfield  {author} {\bibinfo {author} {\bibfnamefont {J.~S.}\ \bibnamefont
  {Bell}},\ }\bibfield  {title} {\bibinfo {title} {On the {E}instein {P}odolsky
  {R}osen paradox},\ }\href
  {https://journals.aps.org/ppf/abstract/10.1103/PhysicsPhysiqueFizika.1.195}
  {\bibfield  {journal} {\bibinfo  {journal} {Phys. Phys. Fiz.}\ }\textbf
  {\bibinfo {volume} {1}},\ \bibinfo {pages} {195} (\bibinfo {year}
  {1964})}\BibitemShut {NoStop}%
\bibitem [{\citenamefont {Bennett}\ \emph {et~al.}(1993)\citenamefont
  {Bennett}, \citenamefont {Brassard}, \citenamefont {Cr\'epeau}, \citenamefont
  {Jozsa}, \citenamefont {Peres},\ and\ \citenamefont
  {Wootters}}]{bennett1993teleporting}%
  \BibitemOpen
  \bibfield  {author} {\bibinfo {author} {\bibfnamefont {C.~H.}\ \bibnamefont
  {Bennett}}, \bibinfo {author} {\bibfnamefont {G.}~\bibnamefont {Brassard}},
  \bibinfo {author} {\bibfnamefont {C.}~\bibnamefont {Cr\'epeau}}, \bibinfo
  {author} {\bibfnamefont {R.}~\bibnamefont {Jozsa}}, \bibinfo {author}
  {\bibfnamefont {A.}~\bibnamefont {Peres}},\ and\ \bibinfo {author}
  {\bibfnamefont {W.~K.}\ \bibnamefont {Wootters}},\ }\bibfield  {title}
  {\bibinfo {title} {Teleporting an unknown quantum state via dual classical
  and {E}instein-{P}odolsky-{R}osen channels},\ }\href
  {https://doi.org/10.1103/PhysRevLett.70.1895} {\bibfield  {journal} {\bibinfo
   {journal} {Phys. Rev. Lett.}\ }\textbf {\bibinfo {volume} {70}},\ \bibinfo
  {pages} {1895} (\bibinfo {year} {1993})}\BibitemShut {NoStop}%
\bibitem [{\citenamefont {Raussendorf}\ \emph {et~al.}(2005)\citenamefont
  {Raussendorf}, \citenamefont {Bravyi},\ and\ \citenamefont
  {Harrington}}]{raussendorf2005long}%
  \BibitemOpen
  \bibfield  {author} {\bibinfo {author} {\bibfnamefont {R.}~\bibnamefont
  {Raussendorf}}, \bibinfo {author} {\bibfnamefont {S.}~\bibnamefont
  {Bravyi}},\ and\ \bibinfo {author} {\bibfnamefont {J.}~\bibnamefont
  {Harrington}},\ }\bibfield  {title} {\bibinfo {title} {Long-range quantum
  entanglement in noisy cluster states},\ }\href
  {https://doi.org/10.1103/PhysRevA.71.062313} {\bibfield  {journal} {\bibinfo
  {journal} {Phys. Rev. A}\ }\textbf {\bibinfo {volume} {71}},\ \bibinfo
  {pages} {062313} (\bibinfo {year} {2005})}\BibitemShut {NoStop}%
\bibitem [{\citenamefont {Tantivasadakarn}\ \emph {et~al.}(2024)\citenamefont
  {Tantivasadakarn}, \citenamefont {Thorngren}, \citenamefont {Vishwanath},\
  and\ \citenamefont {Verresen}}]{tantivasadakarn2024long}%
  \BibitemOpen
  \bibfield  {author} {\bibinfo {author} {\bibfnamefont {N.}~\bibnamefont
  {Tantivasadakarn}}, \bibinfo {author} {\bibfnamefont {R.}~\bibnamefont
  {Thorngren}}, \bibinfo {author} {\bibfnamefont {A.}~\bibnamefont
  {Vishwanath}},\ and\ \bibinfo {author} {\bibfnamefont {R.}~\bibnamefont
  {Verresen}},\ }\bibfield  {title} {\bibinfo {title} {Long-range entanglement
  from measuring symmetry-protected topological phases},\ }\href
  {https://doi.org/10.1103/PhysRevX.14.021040} {\bibfield  {journal} {\bibinfo
  {journal} {Phys. Rev. X}\ }\textbf {\bibinfo {volume} {14}},\ \bibinfo
  {pages} {021040} (\bibinfo {year} {2024})}\BibitemShut {NoStop}%
\bibitem [{\citenamefont {Lu}\ \emph {et~al.}(2022)\citenamefont {Lu},
  \citenamefont {Lessa}, \citenamefont {Kim},\ and\ \citenamefont
  {Hsieh}}]{lu2022measurement}%
  \BibitemOpen
  \bibfield  {author} {\bibinfo {author} {\bibfnamefont {T.-C.}\ \bibnamefont
  {Lu}}, \bibinfo {author} {\bibfnamefont {L.~A.}\ \bibnamefont {Lessa}},
  \bibinfo {author} {\bibfnamefont {I.~H.}\ \bibnamefont {Kim}},\ and\ \bibinfo
  {author} {\bibfnamefont {T.~H.}\ \bibnamefont {Hsieh}},\ }\bibfield  {title}
  {\bibinfo {title} {Measurement as a shortcut to long-range entangled quantum
  matter},\ }\href {https://doi.org/10.1103/PRXQuantum.3.040337} {\bibfield
  {journal} {\bibinfo  {journal} {PRX Quantum}\ }\textbf {\bibinfo {volume}
  {3}},\ \bibinfo {pages} {040337} (\bibinfo {year} {2022})}\BibitemShut
  {NoStop}%
\bibitem [{\citenamefont {Skinner}\ \emph {et~al.}(2019)\citenamefont
  {Skinner}, \citenamefont {Ruhman},\ and\ \citenamefont
  {Nahum}}]{skinner2019measurement}%
  \BibitemOpen
  \bibfield  {author} {\bibinfo {author} {\bibfnamefont {B.}~\bibnamefont
  {Skinner}}, \bibinfo {author} {\bibfnamefont {J.}~\bibnamefont {Ruhman}},\
  and\ \bibinfo {author} {\bibfnamefont {A.}~\bibnamefont {Nahum}},\ }\bibfield
   {title} {\bibinfo {title} {Measurement-induced phase transitions in the
  dynamics of entanglement},\ }\href
  {https://doi.org/10.1103/PhysRevX.9.031009} {\bibfield  {journal} {\bibinfo
  {journal} {Phys. Rev. X}\ }\textbf {\bibinfo {volume} {9}},\ \bibinfo {pages}
  {031009} (\bibinfo {year} {2019})}\BibitemShut {NoStop}%
\bibitem [{\citenamefont {Li}\ \emph {et~al.}(2018)\citenamefont {Li},
  \citenamefont {Chen},\ and\ \citenamefont {Fisher}}]{li2018quantum}%
  \BibitemOpen
  \bibfield  {author} {\bibinfo {author} {\bibfnamefont {Y.}~\bibnamefont
  {Li}}, \bibinfo {author} {\bibfnamefont {X.}~\bibnamefont {Chen}},\ and\
  \bibinfo {author} {\bibfnamefont {M.~P.~A.}\ \bibnamefont {Fisher}},\
  }\bibfield  {title} {\bibinfo {title} {Quantum {Z}eno effect and the
  many-body entanglement transition},\ }\href
  {https://doi.org/10.1103/PhysRevB.98.205136} {\bibfield  {journal} {\bibinfo
  {journal} {Phys. Rev. B}\ }\textbf {\bibinfo {volume} {98}},\ \bibinfo
  {pages} {205136} (\bibinfo {year} {2018})}\BibitemShut {NoStop}%
\bibitem [{\citenamefont {Bao}\ \emph {et~al.}(2020)\citenamefont {Bao},
  \citenamefont {Choi},\ and\ \citenamefont {Altman}}]{bao2020theory}%
  \BibitemOpen
  \bibfield  {author} {\bibinfo {author} {\bibfnamefont {Y.}~\bibnamefont
  {Bao}}, \bibinfo {author} {\bibfnamefont {S.}~\bibnamefont {Choi}},\ and\
  \bibinfo {author} {\bibfnamefont {E.}~\bibnamefont {Altman}},\ }\bibfield
  {title} {\bibinfo {title} {Theory of the phase transition in random unitary
  circuits with measurements},\ }\href
  {https://doi.org/10.1103/PhysRevB.101.104301} {\bibfield  {journal} {\bibinfo
   {journal} {Phys. Rev. B}\ }\textbf {\bibinfo {volume} {101}},\ \bibinfo
  {pages} {104301} (\bibinfo {year} {2020})}\BibitemShut {NoStop}%
\bibitem [{\citenamefont {Jian}\ \emph {et~al.}(2020)\citenamefont {Jian},
  \citenamefont {You}, \citenamefont {Vasseur},\ and\ \citenamefont
  {Ludwig}}]{jian2020measurement}%
  \BibitemOpen
  \bibfield  {author} {\bibinfo {author} {\bibfnamefont {C.-M.}\ \bibnamefont
  {Jian}}, \bibinfo {author} {\bibfnamefont {Y.-Z.}\ \bibnamefont {You}},
  \bibinfo {author} {\bibfnamefont {R.}~\bibnamefont {Vasseur}},\ and\ \bibinfo
  {author} {\bibfnamefont {A.~W.~W.}\ \bibnamefont {Ludwig}},\ }\bibfield
  {title} {\bibinfo {title} {Measurement-induced criticality in random quantum
  circuits},\ }\href {https://doi.org/10.1103/PhysRevB.101.104302} {\bibfield
  {journal} {\bibinfo  {journal} {Phys. Rev. B}\ }\textbf {\bibinfo {volume}
  {101}},\ \bibinfo {pages} {104302} (\bibinfo {year} {2020})}\BibitemShut
  {NoStop}%
\bibitem [{\citenamefont {Garratt}\ \emph {et~al.}(2023)\citenamefont
  {Garratt}, \citenamefont {Weinstein},\ and\ \citenamefont
  {Altman}}]{garratt2023measurements}%
  \BibitemOpen
  \bibfield  {author} {\bibinfo {author} {\bibfnamefont {S.~J.}\ \bibnamefont
  {Garratt}}, \bibinfo {author} {\bibfnamefont {Z.}~\bibnamefont {Weinstein}},\
  and\ \bibinfo {author} {\bibfnamefont {E.}~\bibnamefont {Altman}},\
  }\bibfield  {title} {\bibinfo {title} {Measurements conspire nonlocally to
  restructure critical quantum states},\ }\href
  {https://doi.org/10.1103/PhysRevX.13.021026} {\bibfield  {journal} {\bibinfo
  {journal} {Phys. Rev. X}\ }\textbf {\bibinfo {volume} {13}},\ \bibinfo
  {pages} {021026} (\bibinfo {year} {2023})}\BibitemShut {NoStop}%
\bibitem [{\citenamefont {Nielsen}\ and\ \citenamefont
  {Chuang}(2010)}]{nielsen2010quantum}%
  \BibitemOpen
  \bibfield  {author} {\bibinfo {author} {\bibfnamefont {M.~A.}\ \bibnamefont
  {Nielsen}}\ and\ \bibinfo {author} {\bibfnamefont {I.~L.}\ \bibnamefont
  {Chuang}},\ }\href@noop {} {\emph {\bibinfo {title} {Quantum computation and
  quantum information}}}\ (\bibinfo  {publisher} {Cambridge university press},\
  \bibinfo {year} {2010})\BibitemShut {NoStop}%
\bibitem [{\citenamefont {Koh}\ \emph {et~al.}(2023)\citenamefont {Koh},
  \citenamefont {Sun}, \citenamefont {Motta},\ and\ \citenamefont
  {Minnich}}]{koh2023measurement}%
  \BibitemOpen
  \bibfield  {author} {\bibinfo {author} {\bibfnamefont {J.~M.}\ \bibnamefont
  {Koh}}, \bibinfo {author} {\bibfnamefont {S.-N.}\ \bibnamefont {Sun}},
  \bibinfo {author} {\bibfnamefont {M.}~\bibnamefont {Motta}},\ and\ \bibinfo
  {author} {\bibfnamefont {A.~J.}\ \bibnamefont {Minnich}},\ }\bibfield
  {title} {\bibinfo {title} {Measurement-induced entanglement phase transition
  on a superconducting quantum processor with mid-circuit readout},\ }\href
  {https://doi.org/10.1038/s41567-023-02076-6} {\bibfield  {journal} {\bibinfo
  {journal} {Nat. Phys.}\ }\textbf {\bibinfo {volume} {19}},\ \bibinfo {pages}
  {1314} (\bibinfo {year} {2023})}\BibitemShut {NoStop}%
\bibitem [{\citenamefont {Gullans}\ and\ \citenamefont
  {Huse}(2020)}]{gullans2020scalable}%
  \BibitemOpen
  \bibfield  {author} {\bibinfo {author} {\bibfnamefont {M.~J.}\ \bibnamefont
  {Gullans}}\ and\ \bibinfo {author} {\bibfnamefont {D.~A.}\ \bibnamefont
  {Huse}},\ }\bibfield  {title} {\bibinfo {title} {Scalable probes of
  measurement-induced criticality},\ }\href
  {https://doi.org/10.1103/PhysRevLett.125.070606} {\bibfield  {journal}
  {\bibinfo  {journal} {Phys. Rev. Lett.}\ }\textbf {\bibinfo {volume} {125}},\
  \bibinfo {pages} {070606} (\bibinfo {year} {2020})}\BibitemShut {NoStop}%
\bibitem [{\citenamefont {Noel}\ \emph {et~al.}(2022)\citenamefont {Noel},
  \citenamefont {Niroula}, \citenamefont {Zhu}, \citenamefont {Risinger},
  \citenamefont {Egan}, \citenamefont {Biswas}, \citenamefont {Cetina},
  \citenamefont {Gorshkov}, \citenamefont {Gullans}, \citenamefont {Huse} \emph
  {et~al.}}]{noel2022measurement}%
  \BibitemOpen
  \bibfield  {author} {\bibinfo {author} {\bibfnamefont {C.}~\bibnamefont
  {Noel}}, \bibinfo {author} {\bibfnamefont {P.}~\bibnamefont {Niroula}},
  \bibinfo {author} {\bibfnamefont {D.}~\bibnamefont {Zhu}}, \bibinfo {author}
  {\bibfnamefont {A.}~\bibnamefont {Risinger}}, \bibinfo {author}
  {\bibfnamefont {L.}~\bibnamefont {Egan}}, \bibinfo {author} {\bibfnamefont
  {D.}~\bibnamefont {Biswas}}, \bibinfo {author} {\bibfnamefont
  {M.}~\bibnamefont {Cetina}}, \bibinfo {author} {\bibfnamefont {A.~V.}\
  \bibnamefont {Gorshkov}}, \bibinfo {author} {\bibfnamefont {M.~J.}\
  \bibnamefont {Gullans}}, \bibinfo {author} {\bibfnamefont {D.~A.}\
  \bibnamefont {Huse}}, \emph {et~al.},\ }\bibfield  {title} {\bibinfo {title}
  {Measurement-induced quantum phases realized in a trapped-ion quantum
  computer},\ }\href {https://doi.org/10.1038/s41567-022-01619-7} {\bibfield
  {journal} {\bibinfo  {journal} {Nat. Phys.}\ }\textbf {\bibinfo {volume}
  {18}},\ \bibinfo {pages} {760} (\bibinfo {year} {2022})}\BibitemShut
  {NoStop}%
\bibitem [{\citenamefont {Agrawal}\ \emph {et~al.}(2024)\citenamefont
  {Agrawal}, \citenamefont {Lopez-Piqueres}, \citenamefont {Vasseur},
  \citenamefont {Gopalakrishnan},\ and\ \citenamefont
  {Potter}}]{agrawal2024observing}%
  \BibitemOpen
  \bibfield  {author} {\bibinfo {author} {\bibfnamefont {U.}~\bibnamefont
  {Agrawal}}, \bibinfo {author} {\bibfnamefont {J.}~\bibnamefont
  {Lopez-Piqueres}}, \bibinfo {author} {\bibfnamefont {R.}~\bibnamefont
  {Vasseur}}, \bibinfo {author} {\bibfnamefont {S.}~\bibnamefont
  {Gopalakrishnan}},\ and\ \bibinfo {author} {\bibfnamefont {A.~C.}\
  \bibnamefont {Potter}},\ }\bibfield  {title} {\bibinfo {title} {Observing
  quantum measurement collapse as a learnability phase transition},\ }\href
  {https://doi.org/10.1103/PhysRevX.14.041012} {\bibfield  {journal} {\bibinfo
  {journal} {Phys. Rev. X}\ }\textbf {\bibinfo {volume} {14}},\ \bibinfo
  {pages} {041012} (\bibinfo {year} {2024})}\BibitemShut {NoStop}%
\bibitem [{\citenamefont {{Google AI Quantum and
  Collaborators}}(2023{\natexlab{a}})}]{google2023measurement}%
  \BibitemOpen
  \bibfield  {author} {\bibinfo {author} {\bibnamefont {{Google AI Quantum and
  Collaborators}}},\ }\bibfield  {title} {\bibinfo {title} {Measurement-induced
  entanglement and teleportation on a noisy quantum processor},\ }\href
  {https://doi.org/https://doi.org/10.1038/s41586-023-06505-7} {\bibfield
  {journal} {\bibinfo  {journal} {Nature}\ }\textbf {\bibinfo {volume} {622}},\
  \bibinfo {pages} {481} (\bibinfo {year} {2023}{\natexlab{a}})}\BibitemShut
  {NoStop}%
\bibitem [{\citenamefont {Kamakari}\ \emph {et~al.}(2025)\citenamefont
  {Kamakari}, \citenamefont {Sun}, \citenamefont {Li}, \citenamefont {Thio},
  \citenamefont {Gujarati}, \citenamefont {Fisher}, \citenamefont {Motta},\
  and\ \citenamefont {Minnich}}]{kamakari2025experimental}%
  \BibitemOpen
  \bibfield  {author} {\bibinfo {author} {\bibfnamefont {H.}~\bibnamefont
  {Kamakari}}, \bibinfo {author} {\bibfnamefont {J.}~\bibnamefont {Sun}},
  \bibinfo {author} {\bibfnamefont {Y.}~\bibnamefont {Li}}, \bibinfo {author}
  {\bibfnamefont {J.~J.}\ \bibnamefont {Thio}}, \bibinfo {author}
  {\bibfnamefont {T.~P.}\ \bibnamefont {Gujarati}}, \bibinfo {author}
  {\bibfnamefont {M.~P.~A.}\ \bibnamefont {Fisher}}, \bibinfo {author}
  {\bibfnamefont {M.}~\bibnamefont {Motta}},\ and\ \bibinfo {author}
  {\bibfnamefont {A.~J.}\ \bibnamefont {Minnich}},\ }\bibfield  {title}
  {\bibinfo {title} {Experimental demonstration of scalable cross-entropy
  benchmarking to detect measurement-induced phase transitions on a
  superconducting quantum processor},\ }\href
  {https://doi.org/10.1103/PhysRevLett.134.120401} {\bibfield  {journal}
  {\bibinfo  {journal} {Phys. Rev. Lett.}\ }\textbf {\bibinfo {volume} {134}},\
  \bibinfo {pages} {120401} (\bibinfo {year} {2025})}\BibitemShut {NoStop}%
\bibitem [{\citenamefont {Garratt}\ and\ \citenamefont
  {Altman}(2024)}]{garratt2024probing}%
  \BibitemOpen
  \bibfield  {author} {\bibinfo {author} {\bibfnamefont {S.~J.}\ \bibnamefont
  {Garratt}}\ and\ \bibinfo {author} {\bibfnamefont {E.}~\bibnamefont
  {Altman}},\ }\bibfield  {title} {\bibinfo {title} {Probing postmeasurement
  entanglement without postselection},\ }\href
  {https://doi.org/10.1103/PRXQuantum.5.030311} {\bibfield  {journal} {\bibinfo
   {journal} {PRX Quantum}\ }\textbf {\bibinfo {volume} {5}},\ \bibinfo {pages}
  {030311} (\bibinfo {year} {2024})}\BibitemShut {NoStop}%
\bibitem [{\citenamefont {McGinley}(2024)}]{mcginley2024postselection}%
  \BibitemOpen
  \bibfield  {author} {\bibinfo {author} {\bibfnamefont {M.}~\bibnamefont
  {McGinley}},\ }\bibfield  {title} {\bibinfo {title} {Postselection-free
  learning of measurement-induced quantum dynamics},\ }\href
  {https://doi.org/10.1103/PRXQuantum.5.020347} {\bibfield  {journal} {\bibinfo
   {journal} {PRX Quantum}\ }\textbf {\bibinfo {volume} {5}},\ \bibinfo {pages}
  {020347} (\bibinfo {year} {2024})}\BibitemShut {NoStop}%
\bibitem [{\citenamefont {Raussendorf}\ and\ \citenamefont
  {Briegel}(2001)}]{raussendorf2001one}%
  \BibitemOpen
  \bibfield  {author} {\bibinfo {author} {\bibfnamefont {R.}~\bibnamefont
  {Raussendorf}}\ and\ \bibinfo {author} {\bibfnamefont {H.~J.}\ \bibnamefont
  {Briegel}},\ }\bibfield  {title} {\bibinfo {title} {A one-way quantum
  computer},\ }\href {https://doi.org/10.1103/PhysRevLett.86.5188} {\bibfield
  {journal} {\bibinfo  {journal} {Phys. Rev. Lett.}\ }\textbf {\bibinfo
  {volume} {86}},\ \bibinfo {pages} {5188} (\bibinfo {year}
  {2001})}\BibitemShut {NoStop}%
\bibitem [{\citenamefont {Arute}\ \emph {et~al.}(2019)\citenamefont {Arute},
  \citenamefont {Arya}, \citenamefont {Babbush}, \citenamefont {Bacon},
  \citenamefont {Bardin}, \citenamefont {Barends}, \citenamefont {Biswas},
  \citenamefont {Boixo}, \citenamefont {Brandao}, \citenamefont {Buell} \emph
  {et~al.}}]{arute2019quantum}%
  \BibitemOpen
  \bibfield  {author} {\bibinfo {author} {\bibfnamefont {F.}~\bibnamefont
  {Arute}}, \bibinfo {author} {\bibfnamefont {K.}~\bibnamefont {Arya}},
  \bibinfo {author} {\bibfnamefont {R.}~\bibnamefont {Babbush}}, \bibinfo
  {author} {\bibfnamefont {D.}~\bibnamefont {Bacon}}, \bibinfo {author}
  {\bibfnamefont {J.~C.}\ \bibnamefont {Bardin}}, \bibinfo {author}
  {\bibfnamefont {R.}~\bibnamefont {Barends}}, \bibinfo {author} {\bibfnamefont
  {R.}~\bibnamefont {Biswas}}, \bibinfo {author} {\bibfnamefont
  {S.}~\bibnamefont {Boixo}}, \bibinfo {author} {\bibfnamefont {F.~G.}\
  \bibnamefont {Brandao}}, \bibinfo {author} {\bibfnamefont {D.~A.}\
  \bibnamefont {Buell}}, \emph {et~al.},\ }\bibfield  {title} {\bibinfo {title}
  {Quantum supremacy using a programmable superconducting processor},\ }\href
  {https://doi.org/10.1038/s41586-019-1666-5} {\bibfield  {journal} {\bibinfo
  {journal} {Nature}\ }\textbf {\bibinfo {volume} {574}},\ \bibinfo {pages}
  {505} (\bibinfo {year} {2019})}\BibitemShut {NoStop}%
\bibitem [{\citenamefont {Acharya}\ \emph {et~al.}(2024)\citenamefont
  {Acharya}, \citenamefont {Abanin}, \citenamefont {Aghababaie-Beni},
  \citenamefont {Aleiner}, \citenamefont {Andersen}, \citenamefont {Ansmann},
  \citenamefont {Arute}, \citenamefont {Arya}, \citenamefont {Asfaw},
  \citenamefont {Astrakhantsev} \emph {et~al.}}]{acharya2024quantum}%
  \BibitemOpen
  \bibfield  {author} {\bibinfo {author} {\bibfnamefont {R.}~\bibnamefont
  {Acharya}}, \bibinfo {author} {\bibfnamefont {D.~A.}\ \bibnamefont {Abanin}},
  \bibinfo {author} {\bibfnamefont {L.}~\bibnamefont {Aghababaie-Beni}},
  \bibinfo {author} {\bibfnamefont {I.}~\bibnamefont {Aleiner}}, \bibinfo
  {author} {\bibfnamefont {T.~I.}\ \bibnamefont {Andersen}}, \bibinfo {author}
  {\bibfnamefont {M.}~\bibnamefont {Ansmann}}, \bibinfo {author} {\bibfnamefont
  {F.}~\bibnamefont {Arute}}, \bibinfo {author} {\bibfnamefont
  {K.}~\bibnamefont {Arya}}, \bibinfo {author} {\bibfnamefont {A.}~\bibnamefont
  {Asfaw}}, \bibinfo {author} {\bibfnamefont {N.}~\bibnamefont
  {Astrakhantsev}}, \emph {et~al.},\ }\bibfield  {title} {\bibinfo {title}
  {Quantum error correction below the surface code threshold},\ }\href
  {https://doi.org/10.1038/s41586-024-08449-y} {\bibfield  {journal} {\bibinfo
  {journal} {Nature}\ } (\bibinfo {year} {2024})}\BibitemShut {NoStop}%
\bibitem [{\citenamefont {Vidal}\ and\ \citenamefont
  {Werner}(2002)}]{vidal2003computable}%
  \BibitemOpen
  \bibfield  {author} {\bibinfo {author} {\bibfnamefont {G.}~\bibnamefont
  {Vidal}}\ and\ \bibinfo {author} {\bibfnamefont {R.~F.}\ \bibnamefont
  {Werner}},\ }\bibfield  {title} {\bibinfo {title} {Computable measure of
  entanglement},\ }\href {https://doi.org/10.1103/PhysRevA.65.032314}
  {\bibfield  {journal} {\bibinfo  {journal} {Phys. Rev. A}\ }\textbf {\bibinfo
  {volume} {65}},\ \bibinfo {pages} {032314} (\bibinfo {year}
  {2002})}\BibitemShut {NoStop}%
\bibitem [{\citenamefont {Bao}\ \emph {et~al.}(2024)\citenamefont {Bao},
  \citenamefont {Block},\ and\ \citenamefont {Altman}}]{bao2024finite}%
  \BibitemOpen
  \bibfield  {author} {\bibinfo {author} {\bibfnamefont {Y.}~\bibnamefont
  {Bao}}, \bibinfo {author} {\bibfnamefont {M.}~\bibnamefont {Block}},\ and\
  \bibinfo {author} {\bibfnamefont {E.}~\bibnamefont {Altman}},\ }\bibfield
  {title} {\bibinfo {title} {Finite-time teleportation phase transition in
  random quantum circuits},\ }\href
  {https://doi.org/10.1103/PhysRevLett.132.030401} {\bibfield  {journal}
  {\bibinfo  {journal} {Phys. Rev. Lett.}\ }\textbf {\bibinfo {volume} {132}},\
  \bibinfo {pages} {030401} (\bibinfo {year} {2024})}\BibitemShut {NoStop}%
\bibitem [{\citenamefont {Dehghani}\ \emph {et~al.}(2023)\citenamefont
  {Dehghani}, \citenamefont {Lavasani}, \citenamefont {Hafezi},\ and\
  \citenamefont {Gullans}}]{dehghani2023neural}%
  \BibitemOpen
  \bibfield  {author} {\bibinfo {author} {\bibfnamefont {H.}~\bibnamefont
  {Dehghani}}, \bibinfo {author} {\bibfnamefont {A.}~\bibnamefont {Lavasani}},
  \bibinfo {author} {\bibfnamefont {M.}~\bibnamefont {Hafezi}},\ and\ \bibinfo
  {author} {\bibfnamefont {M.~J.}\ \bibnamefont {Gullans}},\ }\bibfield
  {title} {\bibinfo {title} {Neural-network decoders for measurement induced
  phase transitions},\ }\href {https://doi.org/10.1038/s41467-023-37902-1}
  {\bibfield  {journal} {\bibinfo  {journal} {Nat. Commun.}\ }\textbf {\bibinfo
  {volume} {14}},\ \bibinfo {pages} {2918} (\bibinfo {year}
  {2023})}\BibitemShut {NoStop}%
\bibitem [{\citenamefont {Ippoliti}\ and\ \citenamefont
  {Khemani}(2024)}]{ippoliti2024learnability}%
  \BibitemOpen
  \bibfield  {author} {\bibinfo {author} {\bibfnamefont {M.}~\bibnamefont
  {Ippoliti}}\ and\ \bibinfo {author} {\bibfnamefont {V.}~\bibnamefont
  {Khemani}},\ }\bibfield  {title} {\bibinfo {title} {Learnability transitions
  in monitored quantum dynamics via eavesdropper's classical shadows},\ }\href
  {https://doi.org/10.1103/PRXQuantum.5.020304} {\bibfield  {journal} {\bibinfo
   {journal} {PRX Quantum}\ }\textbf {\bibinfo {volume} {5}},\ \bibinfo {pages}
  {020304} (\bibinfo {year} {2024})}\BibitemShut {NoStop}%
\bibitem [{\citenamefont {Huang}\ \emph {et~al.}(2020)\citenamefont {Huang},
  \citenamefont {Kueng},\ and\ \citenamefont {Preskill}}]{huang2020predicting}%
  \BibitemOpen
  \bibfield  {author} {\bibinfo {author} {\bibfnamefont {H.-Y.}\ \bibnamefont
  {Huang}}, \bibinfo {author} {\bibfnamefont {R.}~\bibnamefont {Kueng}},\ and\
  \bibinfo {author} {\bibfnamefont {J.}~\bibnamefont {Preskill}},\ }\bibfield
  {title} {\bibinfo {title} {Predicting many properties of a quantum system
  from very few measurements},\ }\href
  {https://doi.org/10.1038/s41567-020-0932-7} {\bibfield  {journal} {\bibinfo
  {journal} {Nat. Phys.}\ }\textbf {\bibinfo {volume} {16}},\ \bibinfo {pages}
  {1050} (\bibinfo {year} {2020})}\BibitemShut {NoStop}%
\bibitem [{\citenamefont {Devlin}\ \emph {et~al.}(2019)\citenamefont {Devlin},
  \citenamefont {Chang}, \citenamefont {Lee},\ and\ \citenamefont
  {Toutanova}}]{devlin-etal-2019-bert}%
  \BibitemOpen
  \bibfield  {author} {\bibinfo {author} {\bibfnamefont {J.}~\bibnamefont
  {Devlin}}, \bibinfo {author} {\bibfnamefont {M.-W.}\ \bibnamefont {Chang}},
  \bibinfo {author} {\bibfnamefont {K.}~\bibnamefont {Lee}},\ and\ \bibinfo
  {author} {\bibfnamefont {K.}~\bibnamefont {Toutanova}},\ }\bibfield  {title}
  {\bibinfo {title} {{BERT}: Pre-training of deep bidirectional transformers
  for language understanding},\ }in\ \href
  {https://doi.org/10.18653/v1/N19-1423} {\emph {\bibinfo {booktitle}
  {Proceedings of the 2019 Conference of the North {A}merican Chapter of the
  Association for Computational Linguistics: Human Language Technologies,
  Volume 1 (Long and Short Papers)}}},\ \bibinfo {editor} {edited by\ \bibinfo
  {editor} {\bibfnamefont {J.}~\bibnamefont {Burstein}}, \bibinfo {editor}
  {\bibfnamefont {C.}~\bibnamefont {Doran}},\ and\ \bibinfo {editor}
  {\bibfnamefont {T.}~\bibnamefont {Solorio}}}\ (\bibinfo  {publisher}
  {Association for Computational Linguistics},\ \bibinfo {address}
  {Minneapolis, Minnesota},\ \bibinfo {year} {2019})\ pp.\ \bibinfo {pages}
  {4171--4186}\BibitemShut {NoStop}%
\bibitem [{\citenamefont {Kim}\ \emph {et~al.}(2025)\citenamefont {Kim},
  \citenamefont {Kumar}, \citenamefont {Zhou}, \citenamefont {Xu},
  \citenamefont {Vasseur},\ and\ \citenamefont {Kim}}]{kim2025learning}%
  \BibitemOpen
  \bibfield  {author} {\bibinfo {author} {\bibfnamefont {H.}~\bibnamefont
  {Kim}}, \bibinfo {author} {\bibfnamefont {A.}~\bibnamefont {Kumar}}, \bibinfo
  {author} {\bibfnamefont {Y.}~\bibnamefont {Zhou}}, \bibinfo {author}
  {\bibfnamefont {Y.}~\bibnamefont {Xu}}, \bibinfo {author} {\bibfnamefont
  {R.}~\bibnamefont {Vasseur}},\ and\ \bibinfo {author} {\bibfnamefont {E.-A.}\
  \bibnamefont {Kim}},\ }\href {https://arxiv.org/abs/2508.15895} {\bibinfo
  {title} {Learning measurement-induced phase transitions using attention}}
  (\bibinfo {year} {2025}),\ \Eprint {https://arxiv.org/abs/2508.15895}
  {arXiv:2508.15895 [quant-ph]} \BibitemShut {NoStop}%
\bibitem [{Note1()}]{Note1}%
  \BibitemOpen
  \bibinfo {note} {The value $\epsilon =0.3$ is chosen as we find that this
  improves our bounds for both one- and two-dimensional arrays across wide
  ranges of parameters.}\BibitemShut {Stop}%
\bibitem [{SI()}]{SI}%
  \BibitemOpen
  \href@noop {} {\bibinfo {title} {Supplemental information}}\BibitemShut
  {NoStop}%
\bibitem [{\citenamefont {Hu}\ \emph {et~al.}(2025)\citenamefont {Hu},
  \citenamefont {Liu}, \citenamefont {Zhang},\ and\ \citenamefont
  {Gao}}]{hu2025local}%
  \BibitemOpen
  \bibfield  {author} {\bibinfo {author} {\bibfnamefont {F.}~\bibnamefont
  {Hu}}, \bibinfo {author} {\bibfnamefont {G.}~\bibnamefont {Liu}}, \bibinfo
  {author} {\bibfnamefont {Y.}~\bibnamefont {Zhang}},\ and\ \bibinfo {author}
  {\bibfnamefont {X.}~\bibnamefont {Gao}},\ }\href
  {https://arxiv.org/abs/2508.06614} {\bibinfo {title} {Local diffusion models
  and phases of data distributions}} (\bibinfo {year} {2025}),\ \Eprint
  {https://arxiv.org/abs/2508.06614} {arXiv:2508.06614 [cs.LG]} \BibitemShut
  {NoStop}%
\bibitem [{\citenamefont {Kumar}\ \emph {et~al.}(2026)\citenamefont {Kumar},
  \citenamefont {Zou}, \citenamefont {Negari}, \citenamefont {Melko},\ and\
  \citenamefont {Hsieh}}]{kumar2026unlearnable}%
  \BibitemOpen
  \bibfield  {author} {\bibinfo {author} {\bibfnamefont {T.~A.}\ \bibnamefont
  {Kumar}}, \bibinfo {author} {\bibfnamefont {Y.}~\bibnamefont {Zou}}, \bibinfo
  {author} {\bibfnamefont {A.-R.}\ \bibnamefont {Negari}}, \bibinfo {author}
  {\bibfnamefont {R.~G.}\ \bibnamefont {Melko}},\ and\ \bibinfo {author}
  {\bibfnamefont {T.~H.}\ \bibnamefont {Hsieh}},\ }\href
  {https://arxiv.org/abs/2602.11262} {\bibinfo {title} {Unlearnable phases of
  matter}} (\bibinfo {year} {2026}),\ \Eprint
  {https://arxiv.org/abs/2602.11262} {arXiv:2602.11262 [cond-mat.dis-nn]}
  \BibitemShut {NoStop}%
\bibitem [{\citenamefont {Shalev-Shwartz}\ \emph {et~al.}(2017)\citenamefont
  {Shalev-Shwartz}, \citenamefont {Shamir},\ and\ \citenamefont
  {Shammah}}]{shalevschwartz2017failures}%
  \BibitemOpen
  \bibfield  {author} {\bibinfo {author} {\bibfnamefont {S.}~\bibnamefont
  {Shalev-Shwartz}}, \bibinfo {author} {\bibfnamefont {O.}~\bibnamefont
  {Shamir}},\ and\ \bibinfo {author} {\bibfnamefont {S.}~\bibnamefont
  {Shammah}},\ }\bibfield  {title} {\bibinfo {title} {Failures of
  gradient-based deep learning},\ }in\ \href
  {https://proceedings.mlr.press/v70/shalev-shwartz17a.html} {\emph {\bibinfo
  {booktitle} {Proceedings of the 34th International Conference on Machine
  Learning}}},\ \bibinfo {series} {Proceedings of Machine Learning Research},
  Vol.~\bibinfo {volume} {70},\ \bibinfo {editor} {edited by\ \bibinfo {editor}
  {\bibfnamefont {D.}~\bibnamefont {Precup}}\ and\ \bibinfo {editor}
  {\bibfnamefont {Y.~W.}\ \bibnamefont {Teh}}}\ (\bibinfo  {publisher} {PMLR},\
  \bibinfo {year} {2017})\ pp.\ \bibinfo {pages} {3067--3075}\BibitemShut
  {NoStop}%
\bibitem [{\citenamefont {Napp}\ \emph {et~al.}(2022)\citenamefont {Napp},
  \citenamefont {La~Placa}, \citenamefont {Dalzell}, \citenamefont
  {Brand\~ao},\ and\ \citenamefont {Harrow}}]{napp2022efficient}%
  \BibitemOpen
  \bibfield  {author} {\bibinfo {author} {\bibfnamefont {J.~C.}\ \bibnamefont
  {Napp}}, \bibinfo {author} {\bibfnamefont {R.~L.}\ \bibnamefont {La~Placa}},
  \bibinfo {author} {\bibfnamefont {A.~M.}\ \bibnamefont {Dalzell}}, \bibinfo
  {author} {\bibfnamefont {F.~G. S.~L.}\ \bibnamefont {Brand\~ao}},\ and\
  \bibinfo {author} {\bibfnamefont {A.~W.}\ \bibnamefont {Harrow}},\ }\bibfield
   {title} {\bibinfo {title} {Efficient classical simulation of random shallow
  2d quantum circuits},\ }\href {https://doi.org/10.1103/PhysRevX.12.021021}
  {\bibfield  {journal} {\bibinfo  {journal} {Phys. Rev. X}\ }\textbf {\bibinfo
  {volume} {12}},\ \bibinfo {pages} {021021} (\bibinfo {year}
  {2022})}\BibitemShut {NoStop}%
\bibitem [{\citenamefont {Yang}\ \emph {et~al.}(2024)\citenamefont {Yang},
  \citenamefont {Soleimanifar}, \citenamefont {Bergamaschi},\ and\
  \citenamefont {Preskill}}]{yang2024classical}%
  \BibitemOpen
  \bibfield  {author} {\bibinfo {author} {\bibfnamefont {T.-H.}\ \bibnamefont
  {Yang}}, \bibinfo {author} {\bibfnamefont {M.}~\bibnamefont {Soleimanifar}},
  \bibinfo {author} {\bibfnamefont {T.}~\bibnamefont {Bergamaschi}},\ and\
  \bibinfo {author} {\bibfnamefont {J.}~\bibnamefont {Preskill}},\ }\href
  {https://arxiv.org/abs/2410.23152} {\bibinfo {title} {When can classical
  neural networks represent quantum states?}} (\bibinfo {year} {2024}),\
  \Eprint {https://arxiv.org/abs/2410.23152} {arXiv:2410.23152 [quant-ph]}
  \BibitemShut {NoStop}%
\bibitem [{\citenamefont {Gross}\ and\ \citenamefont
  {Bakr}(2021)}]{gross2021quantum}%
  \BibitemOpen
  \bibfield  {author} {\bibinfo {author} {\bibfnamefont {C.}~\bibnamefont
  {Gross}}\ and\ \bibinfo {author} {\bibfnamefont {W.~S.}\ \bibnamefont
  {Bakr}},\ }\bibfield  {title} {\bibinfo {title} {Quantum gas microscopy for
  single atom and spin detection},\ }\href
  {https://doi.org/10.1038/s41567-021-01370-5} {\bibfield  {journal} {\bibinfo
  {journal} {Nat. Phys.}\ }\textbf {\bibinfo {volume} {17}},\ \bibinfo {pages}
  {1316} (\bibinfo {year} {2021})}\BibitemShut {NoStop}%
\bibitem [{\citenamefont {Shor}(1995)}]{shor1995scheme}%
  \BibitemOpen
  \bibfield  {author} {\bibinfo {author} {\bibfnamefont {P.~W.}\ \bibnamefont
  {Shor}},\ }\bibfield  {title} {\bibinfo {title} {Scheme for reducing
  decoherence in quantum computer memory},\ }\href
  {https://doi.org/10.1103/PhysRevA.52.R2493} {\bibfield  {journal} {\bibinfo
  {journal} {Phys. Rev. A}\ }\textbf {\bibinfo {volume} {52}},\ \bibinfo
  {pages} {R2493} (\bibinfo {year} {1995})}\BibitemShut {NoStop}%
\bibitem [{\citenamefont {Terhal}(2015)}]{terhal2015quantum}%
  \BibitemOpen
  \bibfield  {author} {\bibinfo {author} {\bibfnamefont {B.~M.}\ \bibnamefont
  {Terhal}},\ }\bibfield  {title} {\bibinfo {title} {Quantum error correction
  for quantum memories},\ }\href {https://doi.org/10.1103/RevModPhys.87.307}
  {\bibfield  {journal} {\bibinfo  {journal} {Rev. Mod. Phys.}\ }\textbf
  {\bibinfo {volume} {87}},\ \bibinfo {pages} {307} (\bibinfo {year}
  {2015})}\BibitemShut {NoStop}%
\bibitem [{\citenamefont {Krinner}\ \emph {et~al.}(2022)\citenamefont
  {Krinner}, \citenamefont {Lacroix}, \citenamefont {Remm}, \citenamefont
  {Di~Paolo}, \citenamefont {Genois}, \citenamefont {Leroux}, \citenamefont
  {Hellings}, \citenamefont {Lazar}, \citenamefont {Swiadek}, \citenamefont
  {Herrmann} \emph {et~al.}}]{krinner2022realizing}%
  \BibitemOpen
  \bibfield  {author} {\bibinfo {author} {\bibfnamefont {S.}~\bibnamefont
  {Krinner}}, \bibinfo {author} {\bibfnamefont {N.}~\bibnamefont {Lacroix}},
  \bibinfo {author} {\bibfnamefont {A.}~\bibnamefont {Remm}}, \bibinfo {author}
  {\bibfnamefont {A.}~\bibnamefont {Di~Paolo}}, \bibinfo {author}
  {\bibfnamefont {E.}~\bibnamefont {Genois}}, \bibinfo {author} {\bibfnamefont
  {C.}~\bibnamefont {Leroux}}, \bibinfo {author} {\bibfnamefont
  {C.}~\bibnamefont {Hellings}}, \bibinfo {author} {\bibfnamefont
  {S.}~\bibnamefont {Lazar}}, \bibinfo {author} {\bibfnamefont
  {F.}~\bibnamefont {Swiadek}}, \bibinfo {author} {\bibfnamefont
  {J.}~\bibnamefont {Herrmann}}, \emph {et~al.},\ }\bibfield  {title} {\bibinfo
  {title} {Realizing repeated quantum error correction in a distance-three
  surface code},\ }\href
  {https://doi.org/https://doi.org/10.1038/s41586-022-04566-8} {\bibfield
  {journal} {\bibinfo  {journal} {Nature}\ }\textbf {\bibinfo {volume} {605}},\
  \bibinfo {pages} {669} (\bibinfo {year} {2022})}\BibitemShut {NoStop}%
\bibitem [{\citenamefont {Sivak}\ \emph {et~al.}(2023)\citenamefont {Sivak},
  \citenamefont {Eickbusch}, \citenamefont {Royer}, \citenamefont {Singh},
  \citenamefont {Tsioutsios}, \citenamefont {Ganjam}, \citenamefont {Miano},
  \citenamefont {Brock}, \citenamefont {Ding}, \citenamefont {Frunzio} \emph
  {et~al.}}]{sivak2023real}%
  \BibitemOpen
  \bibfield  {author} {\bibinfo {author} {\bibfnamefont {V.~V.}\ \bibnamefont
  {Sivak}}, \bibinfo {author} {\bibfnamefont {A.}~\bibnamefont {Eickbusch}},
  \bibinfo {author} {\bibfnamefont {B.}~\bibnamefont {Royer}}, \bibinfo
  {author} {\bibfnamefont {S.}~\bibnamefont {Singh}}, \bibinfo {author}
  {\bibfnamefont {I.}~\bibnamefont {Tsioutsios}}, \bibinfo {author}
  {\bibfnamefont {S.}~\bibnamefont {Ganjam}}, \bibinfo {author} {\bibfnamefont
  {A.}~\bibnamefont {Miano}}, \bibinfo {author} {\bibfnamefont
  {B.}~\bibnamefont {Brock}}, \bibinfo {author} {\bibfnamefont
  {A.}~\bibnamefont {Ding}}, \bibinfo {author} {\bibfnamefont {L.}~\bibnamefont
  {Frunzio}}, \emph {et~al.},\ }\bibfield  {title} {\bibinfo {title} {Real-time
  quantum error correction beyond break-even},\ }\href
  {https://doi.org/10.1038/s41586-023-05782-6} {\bibfield  {journal} {\bibinfo
  {journal} {Nature}\ }\textbf {\bibinfo {volume} {616}},\ \bibinfo {pages}
  {50} (\bibinfo {year} {2023})}\BibitemShut {NoStop}%
\bibitem [{\citenamefont {{Google AI Quantum and
  Collaborators}}(2023{\natexlab{b}})}]{google2023suppressing}%
  \BibitemOpen
  \bibfield  {author} {\bibinfo {author} {\bibnamefont {{Google AI Quantum and
  Collaborators}}},\ }\bibfield  {title} {\bibinfo {title} {Suppressing quantum
  errors by scaling a surface code logical qubit},\ }\href
  {https://doi.org/10.1038/s41586-022-05434-1} {\bibfield  {journal} {\bibinfo
  {journal} {Nature}\ }\textbf {\bibinfo {volume} {614}},\ \bibinfo {pages}
  {676} (\bibinfo {year} {2023}{\natexlab{b}})}\BibitemShut {NoStop}%
\bibitem [{\citenamefont {Bluvstein}\ \emph {et~al.}(2024)\citenamefont
  {Bluvstein}, \citenamefont {Evered}, \citenamefont {Geim}, \citenamefont
  {Li}, \citenamefont {Zhou}, \citenamefont {Manovitz}, \citenamefont {Ebadi},
  \citenamefont {Cain}, \citenamefont {Kalinowski}, \citenamefont {Hangleiter}
  \emph {et~al.}}]{bluvstein2024logical}%
  \BibitemOpen
  \bibfield  {author} {\bibinfo {author} {\bibfnamefont {D.}~\bibnamefont
  {Bluvstein}}, \bibinfo {author} {\bibfnamefont {S.~J.}\ \bibnamefont
  {Evered}}, \bibinfo {author} {\bibfnamefont {A.~A.}\ \bibnamefont {Geim}},
  \bibinfo {author} {\bibfnamefont {S.~H.}\ \bibnamefont {Li}}, \bibinfo
  {author} {\bibfnamefont {H.}~\bibnamefont {Zhou}}, \bibinfo {author}
  {\bibfnamefont {T.}~\bibnamefont {Manovitz}}, \bibinfo {author}
  {\bibfnamefont {S.}~\bibnamefont {Ebadi}}, \bibinfo {author} {\bibfnamefont
  {M.}~\bibnamefont {Cain}}, \bibinfo {author} {\bibfnamefont {M.}~\bibnamefont
  {Kalinowski}}, \bibinfo {author} {\bibfnamefont {D.}~\bibnamefont
  {Hangleiter}}, \emph {et~al.},\ }\bibfield  {title} {\bibinfo {title}
  {Logical quantum processor based on reconfigurable atom arrays},\ }\href
  {https://doi.org/10.1038/s41586-023-06927-3} {\bibfield  {journal} {\bibinfo
  {journal} {Nature}\ }\textbf {\bibinfo {volume} {626}},\ \bibinfo {pages}
  {58} (\bibinfo {year} {2024})}\BibitemShut {NoStop}%
\bibitem [{\citenamefont {Iqbal}\ \emph
  {et~al.}(2024{\natexlab{a}})\citenamefont {Iqbal}, \citenamefont
  {Tantivasadakarn}, \citenamefont {Gatterman}, \citenamefont {Gerber},
  \citenamefont {Gilmore}, \citenamefont {Gresh}, \citenamefont {Hankin},
  \citenamefont {Hewitt}, \citenamefont {Horst}, \citenamefont {Matheny} \emph
  {et~al.}}]{iqbal2024topological}%
  \BibitemOpen
  \bibfield  {author} {\bibinfo {author} {\bibfnamefont {M.}~\bibnamefont
  {Iqbal}}, \bibinfo {author} {\bibfnamefont {N.}~\bibnamefont
  {Tantivasadakarn}}, \bibinfo {author} {\bibfnamefont {T.~M.}\ \bibnamefont
  {Gatterman}}, \bibinfo {author} {\bibfnamefont {J.~A.}\ \bibnamefont
  {Gerber}}, \bibinfo {author} {\bibfnamefont {K.}~\bibnamefont {Gilmore}},
  \bibinfo {author} {\bibfnamefont {D.}~\bibnamefont {Gresh}}, \bibinfo
  {author} {\bibfnamefont {A.}~\bibnamefont {Hankin}}, \bibinfo {author}
  {\bibfnamefont {N.}~\bibnamefont {Hewitt}}, \bibinfo {author} {\bibfnamefont
  {C.~V.}\ \bibnamefont {Horst}}, \bibinfo {author} {\bibfnamefont
  {M.}~\bibnamefont {Matheny}}, \emph {et~al.},\ }\bibfield  {title} {\bibinfo
  {title} {Topological order from measurements and feed-forward on a trapped
  ion quantum computer},\ }\href {https://doi.org/10.1038/s42005-024-01698-3}
  {\bibfield  {journal} {\bibinfo  {journal} {Commun. Phys.}\ }\textbf
  {\bibinfo {volume} {7}},\ \bibinfo {pages} {205} (\bibinfo {year}
  {2024}{\natexlab{a}})}\BibitemShut {NoStop}%
\bibitem [{\citenamefont {Iqbal}\ \emph
  {et~al.}(2024{\natexlab{b}})\citenamefont {Iqbal}, \citenamefont
  {Tantivasadakarn}, \citenamefont {Verresen}, \citenamefont {Campbell},
  \citenamefont {Dreiling}, \citenamefont {Figgatt}, \citenamefont {Gaebler},
  \citenamefont {Johansen}, \citenamefont {Mills}, \citenamefont {Moses} \emph
  {et~al.}}]{iqbal2024non}%
  \BibitemOpen
  \bibfield  {author} {\bibinfo {author} {\bibfnamefont {M.}~\bibnamefont
  {Iqbal}}, \bibinfo {author} {\bibfnamefont {N.}~\bibnamefont
  {Tantivasadakarn}}, \bibinfo {author} {\bibfnamefont {R.}~\bibnamefont
  {Verresen}}, \bibinfo {author} {\bibfnamefont {S.~L.}\ \bibnamefont
  {Campbell}}, \bibinfo {author} {\bibfnamefont {J.~M.}\ \bibnamefont
  {Dreiling}}, \bibinfo {author} {\bibfnamefont {C.}~\bibnamefont {Figgatt}},
  \bibinfo {author} {\bibfnamefont {J.~P.}\ \bibnamefont {Gaebler}}, \bibinfo
  {author} {\bibfnamefont {J.}~\bibnamefont {Johansen}}, \bibinfo {author}
  {\bibfnamefont {M.}~\bibnamefont {Mills}}, \bibinfo {author} {\bibfnamefont
  {S.~A.}\ \bibnamefont {Moses}}, \emph {et~al.},\ }\bibfield  {title}
  {\bibinfo {title} {Non-abelian topological order and anyons on a trapped-ion
  processor},\ }\href {https://doi.org/10.1038/s41586-023-06934-4} {\bibfield
  {journal} {\bibinfo  {journal} {Nature}\ }\textbf {\bibinfo {volume} {626}},\
  \bibinfo {pages} {505} (\bibinfo {year} {2024}{\natexlab{b}})}\BibitemShut
  {NoStop}%
\bibitem [{\citenamefont {Xu}\ \emph {et~al.}(2024)\citenamefont {Xu},
  \citenamefont {Sun}, \citenamefont {Wang}, \citenamefont {Li}, \citenamefont
  {Zhu}, \citenamefont {Dong}, \citenamefont {Deng}, \citenamefont {Zhang},
  \citenamefont {Chen}, \citenamefont {Wu} \emph {et~al.}}]{xu2024non}%
  \BibitemOpen
  \bibfield  {author} {\bibinfo {author} {\bibfnamefont {S.}~\bibnamefont
  {Xu}}, \bibinfo {author} {\bibfnamefont {Z.-Z.}\ \bibnamefont {Sun}},
  \bibinfo {author} {\bibfnamefont {K.}~\bibnamefont {Wang}}, \bibinfo {author}
  {\bibfnamefont {H.}~\bibnamefont {Li}}, \bibinfo {author} {\bibfnamefont
  {Z.}~\bibnamefont {Zhu}}, \bibinfo {author} {\bibfnamefont {H.}~\bibnamefont
  {Dong}}, \bibinfo {author} {\bibfnamefont {J.}~\bibnamefont {Deng}}, \bibinfo
  {author} {\bibfnamefont {X.}~\bibnamefont {Zhang}}, \bibinfo {author}
  {\bibfnamefont {J.}~\bibnamefont {Chen}}, \bibinfo {author} {\bibfnamefont
  {Y.}~\bibnamefont {Wu}}, \emph {et~al.},\ }\bibfield  {title} {\bibinfo
  {title} {Non-abelian braiding of fibonacci anyons with a superconducting
  processor},\ }\href {https://doi.org/10.1038/s41567-024-02529-6} {\bibfield
  {journal} {\bibinfo  {journal} {Nat. Phys.}\ }\textbf {\bibinfo {volume}
  {20}},\ \bibinfo {pages} {1469} (\bibinfo {year} {2024})}\BibitemShut
  {NoStop}%
\bibitem [{cir()}]{cirqcode}%
  \BibitemOpen
  \href
  {https://github.com/quantumlib/ReCirq/tree/cbda8e53a1f7cdc4b7f1ad8c25eefd10084cc4b2/recirq/cluster_state_mipt}
  {\bibinfo {title} {Re{C}irq}},\ \bibinfo {note} {{G}ithub
  repository}\BibitemShut {NoStop}%
\bibitem [{\citenamefont {Hou}()}]{hou2025mlmipt}%
  \BibitemOpen
  \bibfield  {author} {\bibinfo {author} {\bibfnamefont {W.}~\bibnamefont
  {Hou}},\ }\href {https://github.com/WandaHou/ML-MIPT} {\bibinfo {title}
  {Machine learning the effects of many quantum measurements}},\ \bibinfo
  {note} {{G}ithub repository}\BibitemShut {NoStop}%
\bibitem [{tra(2024)}]{transformers2024}%
  \BibitemOpen
  \href {https://github.com/huggingface/transformers} {\bibinfo {title}
  {Hugging face transformers library}} (\bibinfo {year} {2024}),\ \bibinfo
  {note} {accessed: 2025-05-05}\BibitemShut {NoStop}%
\bibitem [{\citenamefont {Han}\ \emph {et~al.}(2018)\citenamefont {Han},
  \citenamefont {Wang}, \citenamefont {Fan}, \citenamefont {Wang},\ and\
  \citenamefont {Zhang}}]{PhysRevX.8.031012}%
  \BibitemOpen
  \bibfield  {author} {\bibinfo {author} {\bibfnamefont {Z.-Y.}\ \bibnamefont
  {Han}}, \bibinfo {author} {\bibfnamefont {J.}~\bibnamefont {Wang}}, \bibinfo
  {author} {\bibfnamefont {H.}~\bibnamefont {Fan}}, \bibinfo {author}
  {\bibfnamefont {L.}~\bibnamefont {Wang}},\ and\ \bibinfo {author}
  {\bibfnamefont {P.}~\bibnamefont {Zhang}},\ }\bibfield  {title} {\bibinfo
  {title} {Unsupervised generative modeling using matrix product states},\
  }\href {https://doi.org/10.1103/PhysRevX.8.031012} {\bibfield  {journal}
  {\bibinfo  {journal} {Phys. Rev. X}\ }\textbf {\bibinfo {volume} {8}},\
  \bibinfo {pages} {031012} (\bibinfo {year} {2018})}\BibitemShut {NoStop}%
\bibitem [{\citenamefont {{Geng}}\ \emph {et~al.}(2022)\citenamefont {{Geng}},
  \citenamefont {{Hu}},\ and\ \citenamefont {{Zou}}}]{2022MLS&T...3a5020G}%
  \BibitemOpen
  \bibfield  {author} {\bibinfo {author} {\bibfnamefont {C.}~\bibnamefont
  {{Geng}}}, \bibinfo {author} {\bibfnamefont {H.-Y.}\ \bibnamefont {{Hu}}},\
  and\ \bibinfo {author} {\bibfnamefont {Y.}~\bibnamefont {{Zou}}},\ }\bibfield
   {title} {\bibinfo {title} {{Differentiable programming of isometric tensor
  networks}},\ }\href {https://doi.org/10.1088/2632-2153/ac48a2} {\bibfield
  {journal} {\bibinfo  {journal} {Machine Learning: Science and Technology}\
  }\textbf {\bibinfo {volume} {3}},\ \bibinfo {eid} {015020} (\bibinfo {year}
  {2022})},\ \Eprint {https://arxiv.org/abs/2110.03898} {arXiv:2110.03898
  [quant-ph]} \BibitemShut {NoStop}%
\bibitem [{\citenamefont {Schumacher}\ and\ \citenamefont
  {Nielsen}(1996)}]{schumacher1996quantum}%
  \BibitemOpen
  \bibfield  {author} {\bibinfo {author} {\bibfnamefont {B.}~\bibnamefont
  {Schumacher}}\ and\ \bibinfo {author} {\bibfnamefont {M.~A.}\ \bibnamefont
  {Nielsen}},\ }\bibfield  {title} {\bibinfo {title} {Quantum data processing
  and error correction},\ }\href {https://doi.org/10.1103/PhysRevA.54.2629}
  {\bibfield  {journal} {\bibinfo  {journal} {Phys. Rev. A}\ }\textbf {\bibinfo
  {volume} {54}},\ \bibinfo {pages} {2629} (\bibinfo {year}
  {1996})}\BibitemShut {NoStop}%
\end{thebibliography}
\end{document}